\documentclass[fleqn,usenatbib]{mnras}

\usepackage{newtxtext,newtxmath}

\usepackage[T1]{fontenc}
\usepackage{anyfontsize}

\DeclareRobustCommand{\VAN}[3]{#2}
\let\VANthebibliography\thebibliography
\def\thebibliography{\DeclareRobustCommand{\VAN}[3]{##3}\VANthebibliography}

\usepackage{graphicx}	
\usepackage{multirow}
\usepackage{booktabs}

\newcommand{\pev}{\,$\tilde{\bf e}_l$}
\newcommand{\pevl}{\,$\tilde{\bf e}_{l'}$}
\newcommand{\healpix}{\texttt{HEALPix}}
\newcommand{\nside}{\,$N_{\rm side}$}
\newcommand{\planck}{\textit{Planck}}
\newcommand{\smica}{\texttt{SMICA}}
\newcommand{\fsky}{\,$f_{\rm sky}$}
\newcommand{\isap}{\texttt{iSAP}}

\title[CMB low multipole alignments across WMAP and \planck\ data releases]{CMB low multipole alignments across WMAP and \planck\ data releases}

\author[Sanjeet K. Patel et al.]{
Sanjeet Kumar Patel,$^{1}$
Pavan Kumar Aluri,$^{1}$\thanks{E-mail: pavanaluri.phy@iitbhu.ac.in}
John P. Ralston$^{2}$
\\
$^{1}$Department of Physics, Indian Institute of Technology (BHU), Varanasi - 221005, U.P., India\\
$^{2}$Department of Physics {\&} Astronomy, University of Kansas, Lawrence, KS 66045, USA
}

\date{Accepted XXX. Received YYY; in original form ZZZ}

\pubyear{\the\year{}}

\begin{document}
\label{firstpage}
\pagerange{\pageref{firstpage}--\pageref{lastpage}}
\maketitle

\begin{abstract}
The first observations of the cosmic microwave background (CMB) from NASA's \emph{Wilkinson Microwave Anisotropy Probe (WMAP)} led to finding ‘alignment’ anomalies not expected from fluctuations in the isotropic cosmological model. We study the data of all 8 full-sky public releases since then to test for anomalous alignments and shapes of the first 60 multipoles, i.e. over the range $2\leq l \leq 61$. We use rotationally invariant and covariant statistics to test isotropy of all subsequent WMAP data releases, along with those from the ESA’s \planck\ mission. Anomalous alignments among the multipoles $l=1, 2, 3$ are very consistent and robust. More alignments are detected, some of them new, while significance is diluted by the large range of the search. Power entropy, a measure of the randomness of the multipoles, is consistently anomalous at about $2\sigma$ level or better across all data releases. It appears that the CMB is not as random as the cosmological principle predicts on large angular scales. 
\end{abstract}




\section{Introduction}
The cosmological principle is a foundational assumption of modern cosmology. Though merely a simplifying assumption initially, the assumption became a principle when its far-reaching implications became appreciated. A large body of current cosmological data, and particularly observations of the cosmic microwave background (CMB), invite testing the assumption of isotropy. Since isotropy is a symmetry, it is beautifully well defined to enable model-independent tests. Tests began with release of the first year data from NASA’s \emph{Wilkinson Microwave Anisotropy Probe (WMAP)} space mission, when full-sky high-resolution cleaned CMB maps were made publicly available for the first time~\citep{wmap1maps,wmap1fg}.

It is important to know that the CMB ``power'' statistic (conventionally $C_{l}$) is a rotationally invariant number that itself has no information about rotational invariance (\emph{statistical isotropy}) of the data. All directional information is summed out in computing $C_{l}$. As a result other statistics are needed to test the directional features of the CMB. Anomalous alignments among low multipoles of CMB temperature sky were one of the first instances of isotropy violation seen in the CMB sky~\citep{tegmark04,ralston04,copi04}. The phenomena gained significant attention from the cosmology community by potentially indicating a preferred direction for our universe~\citep{schwarz04,bielewicz04,bielewicz05,land05,copi06,abramo06,samal08,samal09,gruppuso10,copi15,pinkwart18,oliveira20}.
As new full-sky CMB maps from WMAP and later from ESA's \planck\ space probe became available, many of the anomalies were found to still persist. See for example~\cite{cmbanomplk2016}, \cite{beyondlcdm2016} and \cite{cp2023} for a review and current status of various instances of isotropy violation seen in cosmological data including CMB. The WMAP and \planck\ collaborations themselves made a dedicated analysis of these anomalies and found them to be at a similar significance~\citep{wmap7anom,plk13anom,plk15anom,plk18anom}. No interpretations are agreed, and some attributed the observations to chance occurrences. Possible correlations between some of these anomalies were explored, for example, in \cite{muir2018}. The current status of some of the anomalies in ($E$-mode) polarized CMB sky vis-a-vis the anomalies originally seen in CMB temperature data were studied, for example, in \cite{ruishi2023}. 

In this paper we are particularly interested in testing statistical isotropy of CMB modes across all full-sky CMB data releases. The maps available include WMAP's one year to (full) nine years of data, and \planck's nominal, full and legacy mission data sets. Unless otherwise specified, a multipole is `anomalous' if the $p$-value of an associated statistic is less than 5\%. We treat all releases uniformly, while determining significance with exhaustive case-by-case simulations for each release. From each data set we study multipoles in the range $2\leq l \leq 61$.
We note that a statistical penalty will exist for conducting extensive random searches. So rather than tuning the range of study, we decided to accept the penalty of reviewing a wide range of multipoles for the sake of uniformity. Our results shed light on the nature and robustness of anomalous features of CMB modes with the accumulation of data over the years. In brief, some statistics give no evidence against the isotropic null hypothesis. Yet too many other statistics too often contradict isotropy, which simply does not fit the data.

\section{Analysis Procedure}
In this section, we outline the methods and statistics used here to probe anisotropy and alignments among CMB multipoles. We primarily use the \emph{Power tensor} (PT) method~\citep{ralston04} that is briefly described below.

\subsection{Power tensor}
CMB anisotropies are akin to signal on a sphere, and are thus expanded in terms of spherical harmonics, $Y_{lm} (\hat{n})$, which form a suitable basis as
\begin{equation}
    \Delta T (\hat{n}) = \sum_{l=2}^\infty \sum_{m=-l}^l a_{lm} Y_{lm} (\hat{n})\,,
\end{equation}
where $\hat{n}=(\theta,\phi)$, being the position coordinates on a sphere. Here $\Delta T(\hat{n})$ represents CMB temperature anisotropies and $a_{lm}$ are the coefficients of expansion in that basis. The contributions of $l=0, \,1$ are conventionally treated separately, while we will return to $l=1$ in another section. 

\emph{Power tensor} `$A_{ij}$' is then defined as the quadratic estimator involving $a_{lm}$'s as
\begin{equation}
    A_{ij}(l) = \frac{1}{l(l+1)(2l+1)}\sum_{mm'm''}a^*_{lm}J^i_{mm'}J^j_{m'm''}a_{lm''}\,,
    \label{eq:pt}
\end{equation}
where $J^i$ and $J^j$ are the $(2l+1)$ dimensional angular momentum operator matrices : $\{J^x,\, J^y,\, J^z\}$. Thus $A_{ij}$ is a $3\times 3$ matrix whose eigenvalues and eigenvectors inform us of the nature of a particular multipole `$l$' we are interested in. The purpose of the PT is to develop directional statistics for every multipole. We refer to the eigenvector corresponding to the largest eigenvalue as the \emph{principal eigenvector} (PEV or PT-PEV) and assign this as an axis of anisotropy for a multiple. When needed, we will use the symbol \pev\, to denote a PEV corresponding to a multipole `$l$'. Note that eigenvectors are headless, meaning that they do not represent a vector with a signed direction, but instead represent an axis.

PT in equation~(\ref{eq:pt}) s defined so that when isotropy holds the expectation value is given by
\begin{equation}
    \langle A_{ij} \rangle = \frac{C_l}{3} \delta_{ij}\,,
    \label{eq:pt-expt}
\end{equation}
where $C_l$ is the underlying angular power spectrum of the CMB temperature anisotropies. Thus, the eigenvalues add up to the total power, $C_l$, in a multipole `$l$'. In terms of its eigenvalues and eigenvectors, PT can thus be written as
\begin{equation}
    A_{ij} = \sum_{\alpha=1}^3 e_i^\alpha \Lambda^\alpha e_j^\alpha\,,
    \label{eq:pt-ev}
\end{equation}
where $\Lambda^\alpha$ and ${\bf e}^\alpha$ ($\alpha=1,2,3$) are the three eigenvalues and eigenvectors respectively, and $i,j=\{1,2,3\}$ denotes components of the eigenvectors, ${\bf e}^\alpha$.

Let us denote the normalized eigenvalues as $\lambda_\alpha = \Lambda_\alpha/(\sum_\beta \Lambda_\beta)$. To characterize anisotropy of a multipole `$l$', we use what is called the \emph{Power entropy} (PE) defined as
\begin{equation}
    S(l) = -\sum_{\alpha=1}^3 \lambda_\alpha \ln(\lambda_\alpha)\,.
\label{eq:pe}    
\end{equation}
This is the same entropy defined by von Neumann and Shannon. When extracted from a density matrix (a normalized positive definite matrix, such as PT), it is the entropy of quantum information theory~\citep{Ballentine2015}. Thus, PE is a true information entropy measuring the randomness of the eigenvalues of PT.

Here are some of its properties: In essence, PT maps the intricate pattern on the sphere corresponding to a multipole `$l$' onto an ellipsoid. The (normalized) eigenvalues of PT tell us how deformed the ellipsoid is, and the direction of the largest eigenvalue can be taken to represent a preferred direction for that multipole. When isotropy holds, all eigenvalues are equal, viz., $\lambda_\alpha=1/3$, as is obvious from equation~(\ref{eq:pt-expt}). This represents a perfectly symmetric case for a multipole that is mapped to a sphere. On the other hand, when a multipole is highly anisotropic, it is mapped to an ellipsoid that is highly deformed (like a prolate spheroid) along its semimajor axis that naturally reveals an axis of anisotropy. Thus, following equation~(\ref{eq:pe}), PE of a CMB mode falls in the range $0\leq S(l) \leq\ln(3)$. reaches the upper limit, the multipole is maximally isotropic with no preferred orientation. The lower limit represents maximal anisotropy indicating a preferred single axis given by a PEV with eigenvalue `1', and a pattern of eigenvalues \{0, 0, 1\}.

The significance of any deviation from isotropy of a multipole is assessed by comparing its PE observed in the data, $S_{\rm obs}(l)$ with the PE distribution from simulations, \{$S_{\rm sim}(l)$\}. Generating simulated data release by release, incorporating the instrument properties appropriately, is a demanding task that will be described later.

\subsection{Alignment statistic for testing low-$l$ alignments}
Independent of the eigenvalues of PT of a particular multipole, PEVs from different multipoles can themselves be used to compare alignments among multipoles. A natural way to do so is to define the quantity `$x_{l l'}=1-\cos(\alpha_{ll'})$' as an alignment statistic where \pev$\cdot$\pevl$= \cos(\alpha_{ll'})$, and $\alpha_{ll'}$ is the angular separation between the PEVs of any two multipoles $l$ and $l'$.  We can replace symbol $x_{l l'} \rightarrow x$ when the context is clear.  

The standard model of cosmology defines the null model. In this model, CMB temperature anisotropies come from an isotropic Gaussian random field on a sphere. The statistic `$x=1-\cos\alpha$' follows a uniform distribution in the range $[0,1]$ in the standard cosmological model. To show this, compute the area of a spherical cap subtended from an arbitrary $z$-axis to a polar angle $\alpha$. The result is then multiplied by a factor of `2' because PEVs are headless, i.e. undirected axial quantities.  

\subsection{Alignment Tensor}
\label{sec:align}
Collective alignments over a range of multipoles are tested using what is called an
\emph{Alignment tensor} (AT) `$X$'. It is constructed from PEVs ($\tilde{\bf e}_l$)of power tensor, whose decomposition is shown in equation~(\ref{eq:pt-ev}). It is defined as~\citep{samal08}, 
\begin{equation}
X_{ij} = \frac{1}{N_l} \sum_{l}^{N_{l}} \tilde{e}^i_l \tilde{e}^j_{l}.
\label{eq:at}
\end{equation}
Here, the sum is over the selected range of multipoles $l=[l_{min}, l_{max}]$, $\tilde{e}^i_l$ are the components of PEVs, and $N_l$ is the number of multipoles in the sum. AT can also be computed for a chosen set of multipoles to find their collective alignment axis.

Let $\zeta_\alpha$ and ${\bf f}_\alpha$ ($\alpha=1,2,3$) be the three (normalized) eigenvalues and eigenvectors of AT, respectively. The eigenvector corresponding to the largest eigenvalue of AT will be referred to as AT-PEV, denoted by $\tilde{\bf f}$. This eigenvector can be taken to represent the collective alignment axis for any set of multipoles.

Information about the eigenvalues of AT is found in the \emph{Alignment entropy} (AE), defined as
\begin{equation}
S_X = -\sum_{\alpha=1}^3 \zeta_\alpha \ln\zeta_\alpha\,.
\label{eq:ae}
\end{equation}
Just as PE, AE also varies in the range $0\leq S_X\leq\ln(3)$ where the lowest value, $0$, indicates perfect alignment of PEVs and the highest value, $\ln(3)\approx 1.1$, implies perfect isotropy. 
When $S_X\rightarrow 0$, then $\tilde{\bf f}$ represents a collective alignment axis that has been repeated with little variation in the ensemble. It is tempting but false to assume that small $S_X$ would be a direct measure of statistical significance of the direction of  $\tilde{\bf f}$. This is because the eigenvalues and eigenvectors of AT are independent. In the null model, the eigenvectors of AT are distributed isotropically independent of the eigenvalues. In a model representing anisotropy, different samples might have a consistent and statistically significant AT-PEV whether the AT entropy is small or not. More discussion of this appears in Section~\ref{sec:collective} where it is applied to data.

\subsection{Comparing multiple statistics}
Much of the study follows a standard procedure of choosing a threshold $p$-value, denoted $\mathbb{P}$, as the criteria to determine when an observed $p$-value of some statistic will be deemed anomalous. (We will postpone criticism of this procedure for later.) We generally choose $\mathbb{P}=0.05$, i.e. probability of random chance occurrence of 5 per cent or less (corresponding to a $2\sigma$ significance or better). Suppose we find $k_*$ anomalous modes with $p\leq\mathbb{P}$ out of $n$ multipoles being analysed. Then, the cumulative binomial probability with respect to this threshold $p$-value is,
\begin{equation}
P(n,k\geq k_*,\mathbb{P}) = \sum_{k=k_*}^n \binom{n}{k} \mathbb{P}^k(1-\mathbb{P})^{n-k}\,.
\label{eq:cumltprob}
\end{equation}
This is a much more refined estimate than assuming that `$k$' instances of an event with a probability $\mathbb{P}$ would occur with probability $\mathbb{P}^{k}$. Its weakness, however, is that an artificial threshold `$\mathbb{P}$' is chosen to begin with, which creates a bias against counting the most significant statistics. Section~\ref{sec:like} discusses a more complete statistic based on the likelihood of the data in the null distribution. Section~\ref{sec:pvaldis} discusses how the distribution of p-values is used in a classic test of the null distribution.
In some cases, we also test an entire distribution of statistics against the corresponding null distribution. If  $P(x)$ is the theoretical probability distribution function of some statistic `$x$', then the corresponding theoretical cumulative probability distribution function (tCDF) is $\mathcal{P}(x)=\int_0^x P(x') dx' .$  Let the empirical cumulative probability distribution function (eCDF) be denoted by $\mathcal{P}_i$. It is obtained by sorting the statistic `$x$' in ascending order and assigning $\mathcal{P}_i = i/N$, where $N$ is the number of elements in the set. Then the {\it Kuiper statistic} `$V$' is used to compare a tCDF with an eCDF.  It is defined as~\citep{cdfstatsbook},
\begin{equation}
V = D_+ + D_-\,,
\label{eq:kuiperstat}
\end{equation}
where $D_+ = \max\{\mathcal{P}_i-\mathcal{P}(x_i)\}$, and
$D_- = \max\{\mathcal{P}(x_i)-\mathcal{P}_{i-1}\}$, each representing the maximum positive deviation of eCDF above the tCDF and maximum negative deviation of eCDF below the tCDF, respectively.

\section{Observational data and Simulations}
In this study, we analyse CMB maps cleaned using internal linear combination (ILC) method from 1, 3, 5, 7, and 9 yr WMAP data, and \smica-cleaned CMB maps from 2015 and 2018 public data releases of \planck\ satellite mission.
We also study cleaned CMB sky derived from \planck's 2013 data, as explained later.
WMAP data are available from Legacy Archive for Microwave Background Data Analysis
(LAMBDA)\footnote{\url{https://lambda.gsfc.nasa.gov/product/wmap/current/}} and the
\planck\ data is available from Planck Legacy Archive
(PLA)\footnote{\url{https://www.cosmos.esa.int/web/planck/pla}}. Results are consistently reported in the order of their release. All results are evaluated over a consistent range $2\leq l \leq 61$, unless otherwise specified. 

All the WMAP's ILC-cleaned CMB maps (WILC maps) are provided at a pixel resolution of 
\healpix\footnote{\url{https://healpix.sourceforge.io/}} \nside=512 whose beam window function is given by a Gaussian smoothing kernel of full width at half-maximum $(FWHM)=1^\circ$ (degree). WILC maps are derived using a cleaning procedure in real (pixel) space by taking linear combinations of multifrequency raw satellite maps such that ``cosmic'' microwave background signal remains untouched while (nearly) eliminating any foregrounds in the resulting cleaned CMB map~\citep{wmap1fg,eriksen04}.
The \planck\ \smica\ derived CMB maps also employ a similar approach, but the cleaning of raw satellite data maps from different frequency channels is performed in multipole space by taking linear combinations of their spherical harmonic coefficients~\citep{smica2003,smica2008}.
\planck\ provided much higher resolution CMB maps, owing to its better detector capabilities at \nside=2048, with a Gaussian beam of $FWHM=5'$ (arcmin).
For the sake of this analysis, all the CMB maps - data and simulations - were generated or downgraded to have \nside=512 and smoothing level given by a Gaussian beam with $FWHM=1^\circ$. More details on these are provided in Appendix~\ref{apdx:data}.

\begin{figure}
\centering
\includegraphics[width=0.45\textwidth]{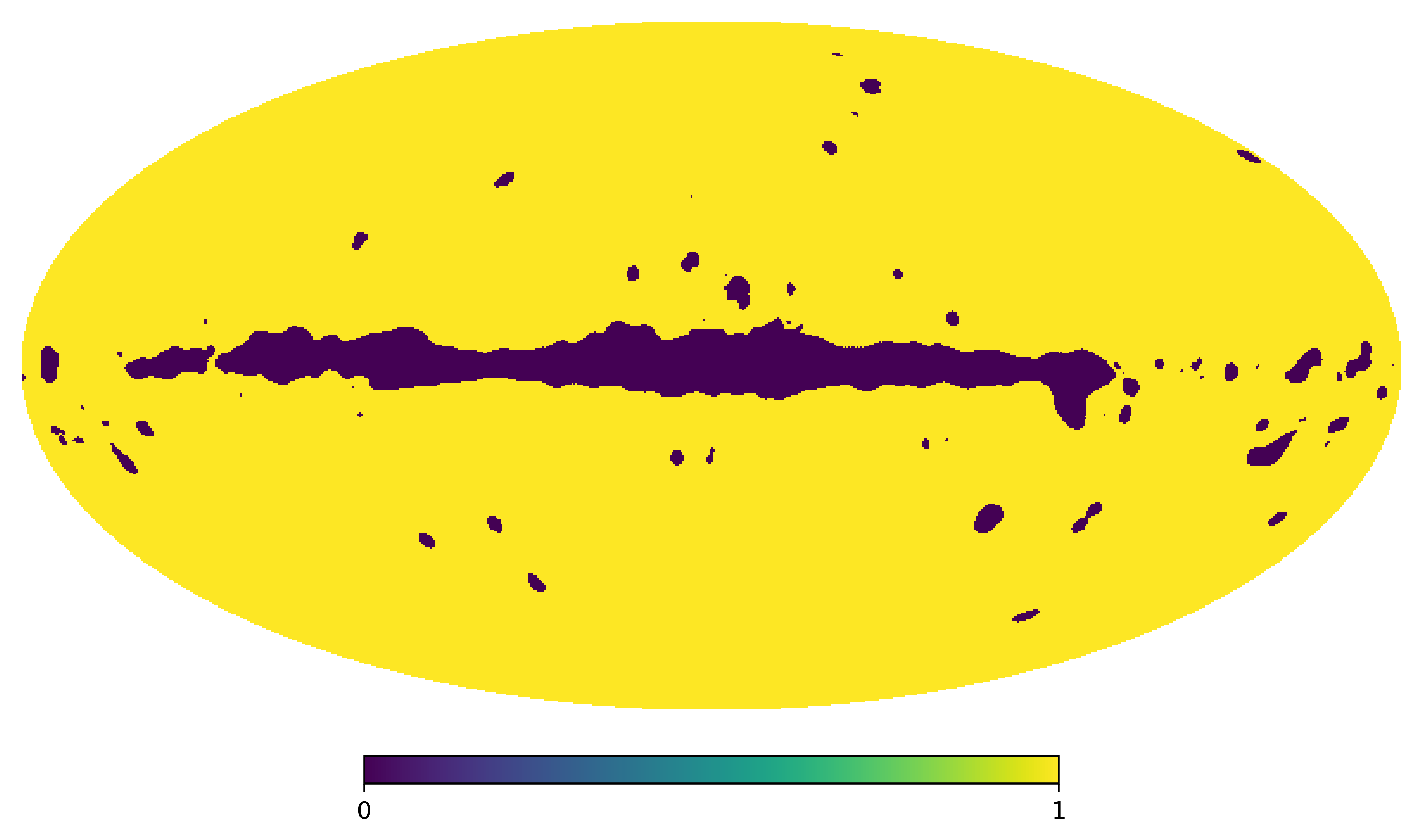}
\caption{Unified mask obtained by merging WMAP’s 9 yr Kp8 mask and \planck's
2018 common inpainting mask, both corresponding to CMB temperature data. This mask has a non-zero sky fraction of \fsky$\approx0.929$.}
\label{fig:our-mask}
\end{figure}

Ideally, any cleaning procedure aims to remove the microwave foregrounds completely. However, in practice, there are still some (visible) foreground residuals present in the \emph{cleaned} CMB maps thus obtained. Such residuals will lead to biased estimates for the quantities of our interest (e.g. statistics being computed) from the cleaned CMB maps. In order to minimize such effects due to galactic residuals, we use a single mask obtained by combining WMAP's nine year Kp8 mask (available at \nside=512) and \planck's 2018 common inpainting mask (provided at \nside=2048).
Since each of these masks is available at different \healpix\ resolutions, they are processed as detailed in Appendix~\ref{apdx:mask} to obtain a unified mask at \healpix\ \nside=512, with
a non-zero sky fraction of \fsky$\approx0.929$.
The mask thus obtained is shown in Fig.~[\ref{fig:our-mask}]. Regions excised by this combination mask are inpainted using the \isap\footnote{\url{http://www.cosmostat.org/software/isap}} ackage, which is a sparsity-based technique to fill those regions of the sky omitted due to masking in a statistically consistent manner with the rest of the available sky.

n order to generate simulations corresponding to WMAP’s cleaned CMB maps, viz. ILC-like CMB maps for various years, we use the weights as published by the WMAP team in the suite of papers accompanying each data release to combine simulated frequency-specific noisy CMB maps~\citep{wmap1fg,wmap3fg,wmap5fg,wmap7fg,wmap9finalmaps}.
The generation and processing of frequency-specific CMB maps with appropriate beam and noise levels, to obtain mock ILC CMB maps for each of the WMAP’s data releases, are explained in Appendix~\ref{apdx:sim}.

Simulated CMB maps corresponding to \planck\ employed \smica\ cleaning procedure are provided by \planck\ collaboration as part of each data release from Planck Legacy Archive. They are referred to as Full Focal Plane (FFP) simulations. We use the FFP \smica\ CMB simulations from 2015 and 2018 data releases as provided by \planck\ collaboration.
In case of \planck\ data release (public release 1/PR1), corresponding simulations are currently not available as they are superseded by Full mission (2015/PR2) and Legacy data (2018/PR3) releases. Hence, we process the \planck\ 2013 data using ILC method in
pixel space, hereafter, referred to as \emph{PR1 ILC}.
The downgrading procedure of \planck\ provided simulated \smica\ CMB maps from 2015 and 2018 data releases, as well as deriving the PR1 ILC CMB map and corresponding simulations are described in Appendix~\ref{apdx:data} and \ref{apdx:sim}.
All these maps are then inpainted using \texttt{iSAP} in the same way as WMAP's ILC maps, using the same mask shown in Fig.~[\ref{fig:our-mask}].

In generating appropriate mock maps to complement ILC-cleaned CMB maps from WMAP’s all data releases and \planck\ PPR1 observations, we found that some care is required in handling the circular beam transfer functions ($b_l$) of low frequency maps in both missions at high-$l$ as provided by respective collaborations. These additional details are presented in Appendix~\ref{apdx:wmap1kmap} and \ref{apdx:plkpr1lfi}.

\section{Results}

\subsection{Anisotropically distributed multipoles}
The current cosmological model based on (statistical) isotropy predicts a null distribution of the eigenvalues of power tensor. To reiterate, the power entropy `$S(l)$'~(equation~\ref{eq:pe}) is a rotationally invariant statistic for each mode `$l$' which summarizes information on the distribution of eigenvalue sizes. We computed $S(l)$ for all the data sets used in this study, and compared the results with the corresponding simulations.

\begin{figure*}
  \centering
  \includegraphics[width=0.95\textwidth]{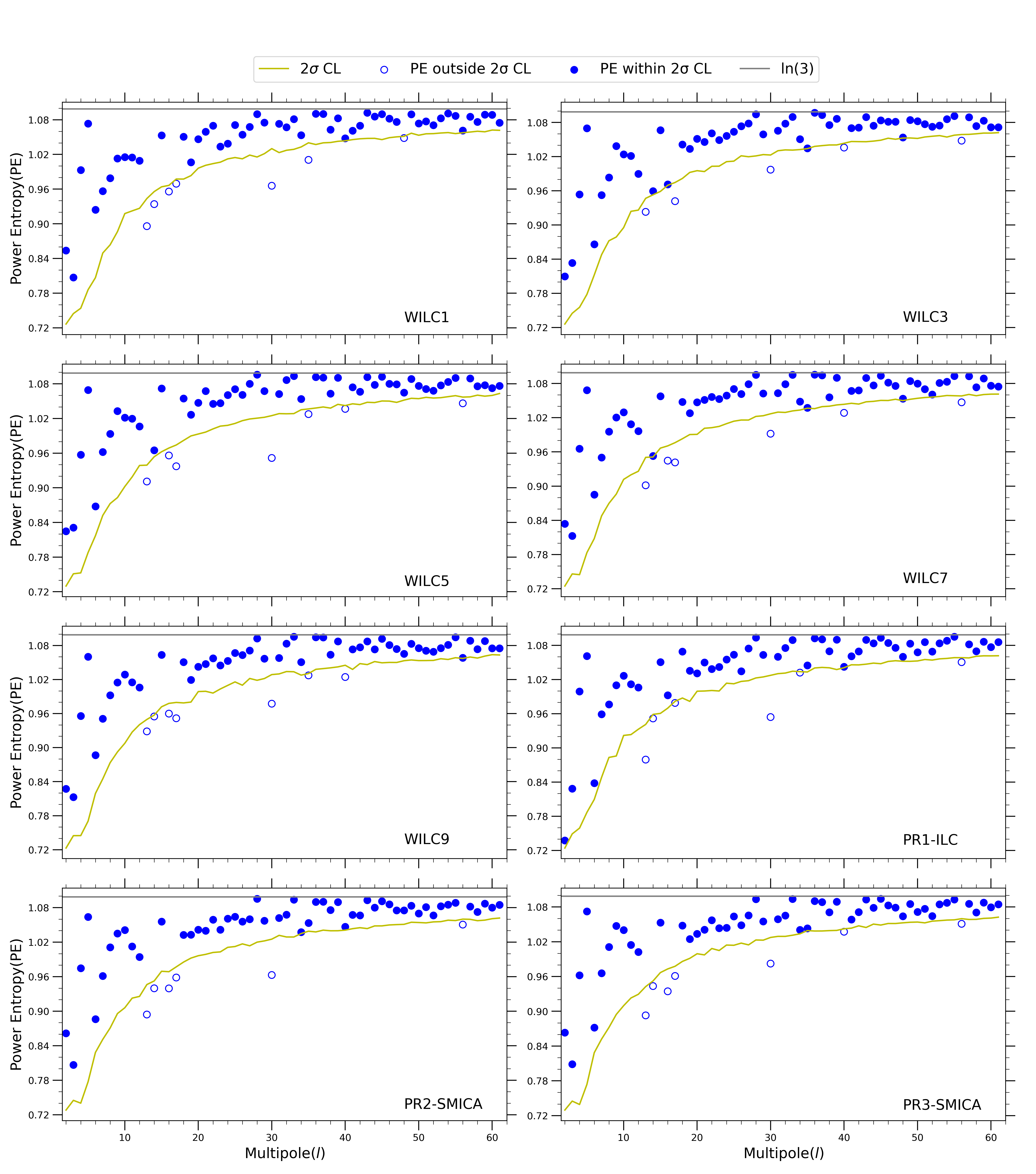}
  \caption{Power entropy of each multipole, $S(l)$ (per equation~\ref{eq:pe})
		of CMB maps from WMAP mission obtained using ILC method on 1, 3, 5, 7, and 9 yr data, as well as from \planck\ PR1 data cleaned using ILC method (\planck\ PR1 ILC), and
		\smica\-cleaned CMB maps from \planck's 2015 and 2018 data releases. The olive green lines delineate $2\sigma$ confidence level (95 per cent) mode by mode derived from simulations.}
\label{fig:pe}
\end{figure*}

Results are shown in Fig.~[\ref{fig:pe}]. The solid grey line at the top of each figure indicates the maximum value $S(l)$ can take, representing a completely isotropic case. The olive green line is the 95 per cent confidence level (CL) for the PE estimated from simulations for each multipole in the range of our interest $l=[2,61]$.

Since early suggestions of anisotropy were found in modes of rather small `$l$', extending to a search over 60 total modes might create a penalty favouring the isotropic null. However, Fig.~[\ref{fig:pe}] shows that modes with anomalous PE are not confined to low $l$, but in some cases appear across the whole range. Open circles show anomalous cases with random chance occurrence probability of $p\leq0.05$. Filled circles have $p>0.05$. Table~\ref{tab:pe-pval} summarizes the anomalous cases.

\begin{table}
\centering
\begin{tabular}{clc}
\hline
Data set & Multipoles, $l$  & Cumulative Bionomial \\
         &                  & Probability \\
\hline
WILC 1yr    & 13, 14, 16, 17, & 0.030\\
            & 30, 35, 48      & \\
WILC 3yr    & 13, 17, 30, 40, & 0.180\\
            & 56              & \\
WILC 5yr    & 13, 16, 17, 30, & 0.030\\
            & 35, 40, 56      & \\
WILC 7yr    & 13, 14, 16, 17, & 0.030\\
            & 30, 40, 56      & \\
WILC 9yr    & 13, 14, 16, 17, & 0.030\\
            & 30, 35, 40      & \\
PR1 ILC     & 13, 14, 17, 30, & 0.079\\
            & 34, 56          & \\
PR2 \smica\ & 13, 14, 16, 17  & 0.079\\
            & 30, 56          & \\
PR3 \smica\ & 13, 14, 16, 17  & 0.030\\
            & 30, 40, 56      & \\
\hline
\end{tabular}
\caption{List of multipoles (second column) whose power entropy has a $p$-value of 5 per cent or less, as found in WILC CMB maps from 1, 3, 5, 7, and 9 yr WMAP data, and \planck\ 2013 data (PR1) cleaned using ILC method, and
the \smica\ CMB maps from \planck\ probe's 2015 and 2018 data releases (mentioned in first column). The last column denotes the cumulative binomial probability to find the observed number of anomalous multipoles or more by random chance in the entire set over the multipole range $l=[2,61]$.}
\label{tab:pe-pval}
\end{table}

From Table~\ref{tab:pe-pval}, we see that the multipoles $l=13$, $17$ and $30$ stand out consistently as anomalous in all of the five ILC maps from WMAP, as well as in \planck's PR1 ILC map, and the \smica\ cleaned CMB maps from PR2 and PR3 data. More anomalous (more rare) entropies are smaller, corresponding to eigenvalues that are further from being equal.

The last column of Table~\ref{tab:pe-pval} indicates the cumulative probability of finding  the number of anomalous modes with $p\leq \mathbb{P}$ as observed in data, given the total number of modes analysed (i.e., $l=2$ to $61$). or example, the cumulative binomial probability (equation~\ref{eq:cumltprob}) of observing 7 items with $p \leq 0.05$ in a random search of 60 cases is $\approx 0.030$. Enough cases are highly significant that the data do not support the isotropic null distribution. Many more multipoles would be deemed anomalous if the threshold $p$-value was chosen to be $\mathbb{P}=0.1$. The large number of cases shown in Fig.~[\ref{fig:pe}] where the observed $p$-values are borderline, or just inside the cut-off curve, indicate that the arbitrary threshold of $\mathbb{P} = 0.05$ creates a bias against detecting anisotropy.

\begin{table}
\centering
\begin{tabular}{c c c}
\hline
Data set &      $\Pi_S$    &  $p$-value\\
\hline
WILC 1yr     &  75.510 & 0.033 \\
WILC 3yr     &  72.779 & 0.067 \\
WILC 5yr     &  74.940 & 0.042 \\
WILC 7yr     &  78.664 & 0.018 \\
WILC 9yr     &  75.603 & 0.028 \\
PR1 ILC      &  78.824 & 0.013 \\
PR2  \smica\ &  74.736 & 0.035 \\
PR3 \smica\  &  77.483 & 0.019 \\
\hline
\end{tabular}
\caption{Values of the \emph{log}-likelihood statistic $\Pi_S$, defined in equation~(\ref{eq:log-like}) based on the null distribution of the power entropy, $S(l)$, are listed in second column for various data sets listed in first column. The statistic values mentioned against different data releases were evaluated using the distribution obtained from a corresponding simulation ensemble of 1000 mocks. The $p$-values of $\Pi_S$ thus obtained are given in the third column. The likelihood-based $p$-values are smaller than those of Table~\ref{tab:pe-pval}, which are evaluated with a cut-off of $p \leq 0.05$. This is evidence that using a hard cut-off on assessing multiple $p$-values tends to create a bias against detecting a signal.}
\label{tab:log-like}
\end{table}

\subsection{A \emph{log}-like statistic}
\label{sec:like}
To deal with a definite bias against rare events caused by choosing a threshold $p$-value `$\mathbb{P}$', we investigated a likelihood-based statistic. Let $f(x_l)$ be the known (simulated) distribution for the statistic `$x_l$' for each multipole `$l$'. We then define a global statistic $\Pi_x$ as
\begin{equation}
\Pi_x = -\sum_{l} \log[1-F(x_l)]\,,
\label{eq:log-like}
\end{equation}
where $F(x_l)$ is the (null) cumulative distribution corresponding to $f(x_l)$. The value of this global statistic in data is then compared to the same global statistic from simulations. By making use of the null distribution of the statistic `$x_l$', there are no arbitrary cut-offs and conclusions are not sensitive to that choice.

In Table~\ref{tab:log-like}, we summarize the results of $\Pi_S$ for the PE statistic, i.e. $x_l=S(l)$, along with the corresponding $p$-values for the multipole range $l=2$ to 61. One can readily see from $p$-value column of Table~\ref{tab:log-like} that the multipoles in this range are collectively anomalous at about $2\sigma$ level or better in almost all data sets. However, using the threshold $\mathbb{P}=0.05$ might cause one to conclude that only few multipoles were anomalous from Table~\ref{tab:pe-pval}. Thus, we conclude that the multipoles, not just few but collectively, indicate an intrinsic anisotropy of the data sets.

For completeness, we report that in three cases (WMAP 1yr, WMAP 7yr, and \planck\ PR1) the 1000-run simulations found no cases of PE lower than the data for  $l=30$. An upper limit of $p=10^{-3}$ was used to make an estimate of the $\Pi_S$ statistic. A very long re-simulation of one of these cases (WILC7) convinced us that this order of magnitude is consistent.

\begin{figure}
\centering
\includegraphics[width=0.98\columnwidth]{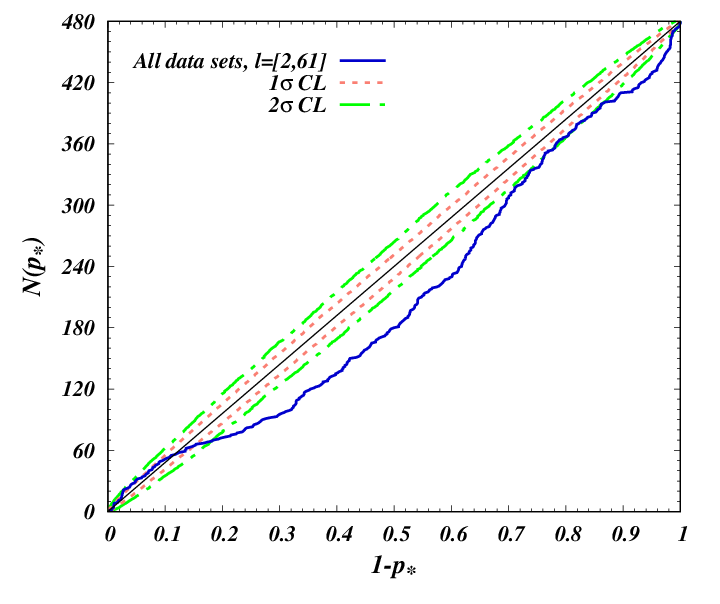}
\caption{ The number of cases of $p<p_{*}$ denoted by `$N(p_{*})$' as a function of `$1-p_{*}$' for the power entropy are plotted as a blue curve. Data from the null distribution tend to be on the black straight line. The $1\sigma$ and $2\sigma$ confidence levels are shown in dashed salmon and dash–dotted green lines, respectively. As discussed in \citet{schweder1982} and Section~\ref{sec:pvaldis}, the deviation of the data from the null around 50 of 480 cases ($y$-axis) indicates that the null model fits about 10 per cent of the data $p$-values.}
\label{fig:PE-PValueCumulative}
\end{figure}

\subsection{$p$-value distributions} 
\label{sec:pvaldis}
An issue in this type of data analysis is that not all statistics and not all data sets are independent. While some of that is controlled by the penalties of the cumulative binomial search over 60 multipoles, a considerable amount of information tends to be wasted. For each study of 60 cosmological multipoles across 8 data releases, we have 480 $480$ $p$-values, of which the vast majority do not pass a test based on significance of $\mathbb{P}=5\%$. Nevertheless, the distribution of these $p$-values has interesting information.

The problem of evaluating a large number of $p$-alues for a given set of data was confronted in the classic paper by~\citet{schweder1982}. In their words, ``We consider a situation in which {\it a large number of tests are made on the same data or are related to the same problem}. A classical example is the one-way layout in the analysis of variance when all pairs of means are compared.'' Italics are ours. The relevance of the method (and the quote) is that no assumptions are made about data elements being independent. The key observation is that the $p$-values of a true null distribution are uniformly distributed on the interval $[0,1]$. If they are not uniformly distributed, it indicates that the null distribution does not fit the data. To illustrate this, \citet{schweder1982} introduced a device called a $p$-value plot. The plot shows the number of cases `$N(p_{*})$' with $p<p_{*}$ as a function of  `$1-p_{*}$', which the null distribution predicts to follow a straight line (if the null is correct).

In Fig.~[\ref{fig:PE-PValueCumulative}], we show such a plot for the 480 $p$-values f PE for 60 multipoles from all the 8 data releases. The first 50 or so cases in the plot near $1-p_{*} \sim 0$ (large $p$-values) are consistent with the null. The remaining points deviate greatly below the line, which means that there are too many small $p$-values to be consistent with the null. The Anderson–Darling (AD) test statistic for the data $p$-value distribution to come from the null distribution is  $10^{-6}$. Keeping in mind that the eight data sets are {\it attempting} to represent the same physical quantities, one might take that with a grain of salt. The result of eight similar but individual tests made on each data release using AD statistic returns a $p$-value of 0.05, 0.09, 0.08, 0.17, 0.09, 0.1, 0.05, and 0.02, in the order listed in the tables. Whether the different data sets are considered to be practically the same, or significantly independent, they repeatedly disfavour the null distribution.

\begin{figure}
\centering
\includegraphics[width=0.98\columnwidth]{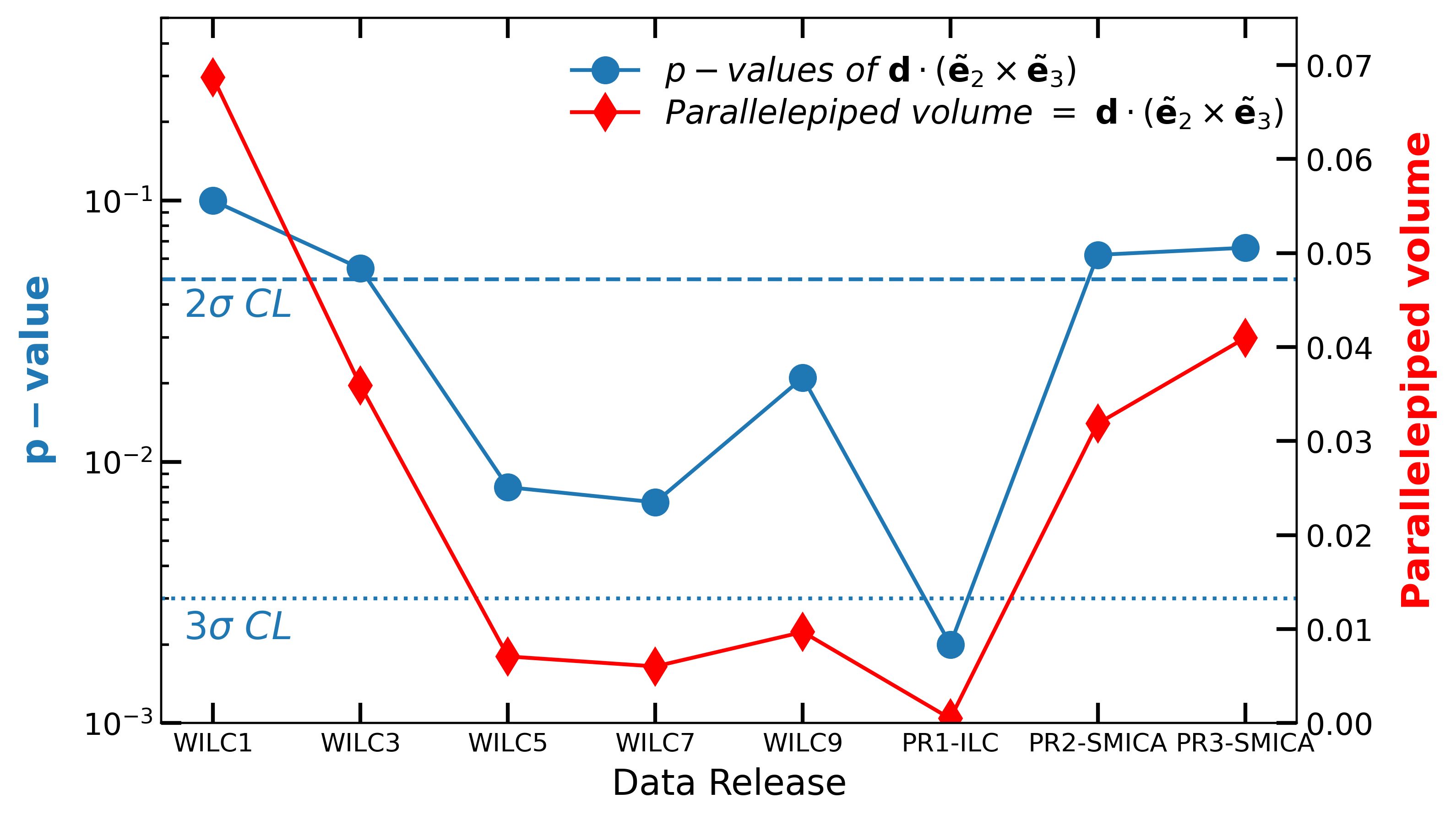}
\caption{$p$-value corresponding to the alignment of CMB dipole, quadrupole, and octopole as quantified by the volume formed by respective axes ${\bf d}$, $\tilde{\bf e}_2$ and
	 $\tilde{\bf e}_3$ are shown as \emph{blue} circles joined by a line using $y1$-axis and the
	 value of the statistic viz. the volume of the parallelepiped formed by these axes $= {\bf d} \cdot (\tilde{\bf e}_2  \times \tilde{\bf e}_3)$
	 is shown using the $y2$-axis with \emph{red} diamond point type joined by a line for all the data releases that we studied. The significance of their alignment is evidently about $2\sigma$ level or better.}
\label{fig:l123-volume}
\end{figure}

\begin{figure*}
  \centering
  \includegraphics[width=0.95\textwidth]{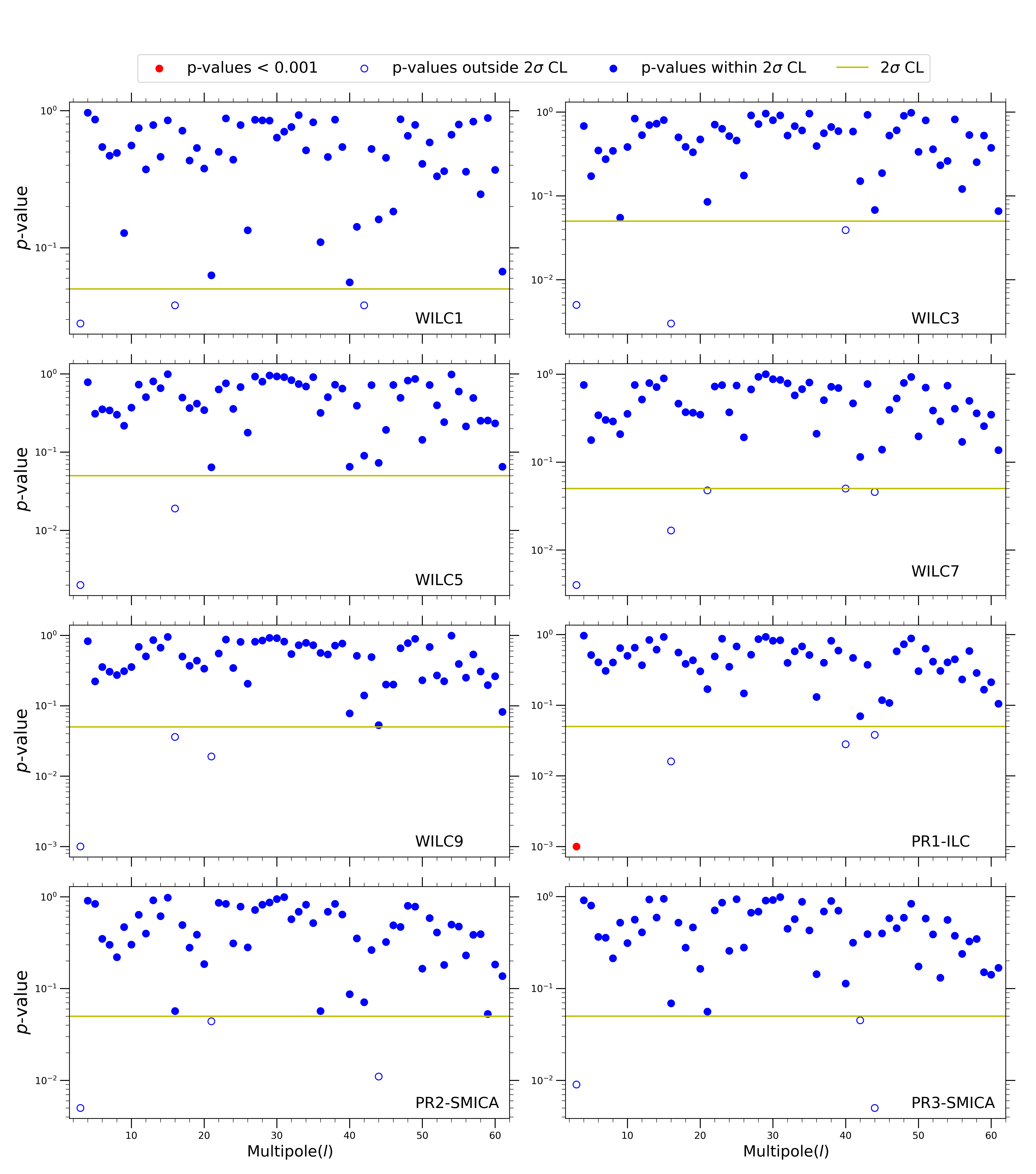}
  \caption{$p$-values corresponding to the alignment of quadrupole with higher order multipoles, i.e. alignment between the said PEVs in each CMB map from various data releases.}
\label{fig:algn-withl2}
\end{figure*}

\subsection{Volume statistic} 
\label{sec:l123vol}
One of the interesting coincidences in the study of CMB low multipole alignments is the close proximity between CMB dipole, quadrupole, and octopole. While the CMB dipole is (predominantly) a consequence of our relative motion with respect to CMB rest frame, and therefore not cosmological, the alignment of $l=2,3$ modes with $l=1$ has gathered significant attention to understand their unexpected correlations (see for example~\cite{schwarz04,ralston04,cmbanomplk2016}). Here, we use a simple statistic, the volume of the parallelepiped formed by their axes, to test their alignment with each other. When they are closely aligned, we have a collapsed parallelepiped with nearly zero volume and large otherwise.

The direction of the CMB dipole is taken to be ${\bf d} = (\ell,b)\approx(264^\circ,48^\circ)$ for the CMB dipole\footnote{\url{https://lambda.gsfc.nasa.gov/education/lambda_graphics/cmb_dipole.html}} in galactic coordinates.
In Fig.~[\ref{fig:l123-volume}], we exhibit the volume of the parallelepiped $={\bf d} \cdot (\tilde{\bf e}_2  \times \tilde{\bf e}_3)$ formed by CMB dipole ($l=1$), and the $l=2,3$ PEVs. It is evident from the figure that these modes are well aligned at about $2\sigma$ level or better depending on specific data release. See for example~\cite{tegmark06}, \cite{bielewicz05}, and \cite{aluri2011} that probed the stability of these large angular modes against foregrounds and varying galactic sky cuts (masks).

\subsection{Multipoles aligned with $l=2$}
Alignment with high significance between the modes $l=2$ and $3$ was initially observed in CMB maps derived from WMAP 1 yr data. Similar alignments were later found among higher multipoles also~\citep{tegmark03,tegmark04,eriksen04,copi04,schwarz04,samal08}.
Interestingly, $l\geq2$ modes were also aligned with the CMB dipole and other preferred axes found in various astronomical data~\citep{ralston04}. Here, we extend the study of alignments with $l=2$ to the entire range of our study. The test statistic is $x:=1-\tilde{\bf e}_l\cdot\tilde{\bf e}_{2} = 1-\cos\alpha_{l2}$', where $\tilde{\bf e}_l$ and $\tilde{\bf e}_{2}$ are the PEVs of multipoles `$l$' and `$2$'.

The results are presented in Fig.~[\ref{fig:algn-withl2}].  A $2\sigma$ anomaly ($p\leq0.05)$ is denoted by an olive green line in each subplot. The octopole mode is anomalously aligned with the quadrupole mode in all CMB maps analysed in this study. Multipoles that are spuriously aligned with the quadrupole are listed in column~2 of Table~\ref{tab:algn-l2-pval}. Third column of Table~\ref{tab:algn-l2-pval} denotes the cumulative null probability of finding the observed number of multipoles or more in the data set. The last column of Table~\ref{tab:algn-l2-pval}, reports the $p$-value of the Kuiper statistic (equation~\ref{eq:kuiperstat}) of the alignment measure `$x$' for the entire set compared to a uniform distribution. Readers can assess the meaning of these results for themselves. The very high significance of early studies discovering the alignments of $l=$ 1, 2, 3 has not changed. When the same data are immersed in a search over 60 multipoles, with $l=1$ excluded, $l=2$ given, and $l=3$ contributing, the estimated significance is greatly reduced.

\begin{table}
\centering
\begin{tabular}{c l c c}
\hline
Data set & Multipoles, $l$ & Cumulative  & $p$-value of $V$\\
         &                 & Probability & \\
\hline
WILC 1yr    & 3, 16, 42      & 0.571 & 0.192\\
WILC 3yr    & 3, 16, 40      & 0.571 & 0.915\\
WILC 5yr    & 3, 16          & 0.801 & 0.796\\
WILC 7yr    & 3, 16, 21, 44  & 0.341 & 0.412\\
WILC 9yr    & 3, 16, 21      & 0.571 & 0.659\\
PR1 ILC     & 3, 16, 40, 44  & 0.341 & 0.116\\
PR2 \smica\ & 3, 21, 44      & 0.571 & 0.531\\
PR3 \smica\ & 3, 42, 44      & 0.571 & 0.673\\
\hline
\end{tabular}
\caption{List of multipoles that are aligned with the quadrupole with a $p$-value $\leq 5$ per cent as found for the eight data releases. The third column denotes the cumulative binomial probability of finding the observed number of anomalous modes or more with $p\leq0.05$. The fourth column gives the probability of the observed Kuiper statistic (V) for the set. The dipole ($l=1$) has been excluded by convention. Including it makes little difference to significance dominated by making 60 searches.}
\label{tab:algn-l2-pval}
\end{table}

Before leaving this section, we report that neglecting the effects of small $p$-values does not seem to be the cause of these results. Repeating the whole analysis with likelihood statistics developed from the simulations (as in Section~\ref{sec:like}) did not find significantly different outcomes.

\subsection{Pairwise alignments among multipoles}
\label{sec:pairwise}
Extensive random searches can dilute statistical significance. That was certainly not a problem in the early history, which was confined to alignments of low multipoles $l$=1, 2, 3. For thoroughness of this study, we nevertheless decided to explore all pairwise combinations of PEVs from all data releases with the statistic `$x:=1-\tilde{\bf e}_l\cdot\tilde{\bf e}_{l'}$'. For the multipole range studied, i.e., $l=[2,61]$, there are $n=60 \times 59/2=1770$ possible combinations to consider for each release. This creates a correspondingly large penalty from the cumulative binomial distribution for whatever anomalies might be found. The results are summarized in Table~\ref{tab:algn-ll-pval}. Nothing of significance is indicated by this search. 
 
\begin{table}
\centering
\begin{tabular}{c c c c}
\hline
Data set &      $k_*$    &  Cumulative  & $p$-value of $V$\\
         & (out of 1770) &  Probability & \\
\hline
WILC 1yr     &  82 & 0.775 & 0.307\\
WILC 3yr     &  92 & 0.366 & 0.562\\
WILC 5yr     &  81 & 0.808 & 0.850\\
WILC 7yr     &  93 & 0.327 & 0.417\\
WILC 9yr     &  81 & 0.808 & 0.548\\
PR1 ILC      &  88 & 0.537 & 0.622\\
PR2  \smica\ &  89 & 0.494 & 0.920\\
PR3 \smica\  &  91 & 0.408 & 0.794\\
\hline
\end{tabular}
\caption{Alignment statistic `$x:=1-\cos\alpha$' for all pairwise combinations ($60\times59/2=1770$ total) for each release. Column 2 lists the number of anomalous alignments ($k_*$) found with $p\leq 0.05$. Column 3 lists the cumulative binomial probabilities to have found $k_*$ modes that are spuriously aligned per $\mathbb{P}=0.05$. Column 4 lists the cumulative Kuiper statistic ($V$) for the search.}
\label{tab:algn-ll-pval}
\end{table}

\subsection{Common set of anomalous multipoles shared by all releases}
\label{sec:commong-anom-l}

Table~\ref{tab:common-modes-pval5pc} lists multipoles that were anomalous at the 95 per cent level in every one of the eight data releases when employing a particular statistic. For example, the PE of $l$= 13, 17, 30 was never observed in any simulation to have a $p$-value exceeding 0.05. (This understates the information: The $p$-values for $l=13$ were 0.016, 0.024, 0,018, 0.016, 0.038, 0.009, 0.014, 0.014. The $l=17$ case is also less likely by chance. For $l =30$ the set was 0.0, 0.017, 0.001, 0.0, 0.004, 0.0, 0.001, 0.001. These $p$-values are in the same order of data releases as mentioned in tables.) 

In brief, we note that the number of multipoles with anomalous PE values is about $2\sigma$ in almost all data releases that we studied. Expected number of alignments that would be outside $2\sigma$ CL are $\sim 3$ when considering alignment between quadrupole and higher order modes that are $59$ in number, and $\sim 88$ rom among all possible pairwise alignments that are $1770$ in number, for the multipole range $l=[2,61]$. We find nearly the same number of multipoles to be aligned with a random chance occurrence probability of $p\leq 0.05$ in both the cases. So, the observed number of alignments are consistent with the expectations, which were further confirmed using a cumulative probability distribution function statistic (Kuiper test).

We also note in passing that though we did not find any collective alignments between $l=2$ and rest of the multipoles or in pairwise alignment tests, a different kind of analysis using the same statistics employed here revealed an interesting pattern. Testing alignments among only even or odd multipoles separately, it was found that the \emph{odd} multipoles from the same multipole range studied here ($l=[2,61]$) have an anomalous alignment entropy (equation~\ref{eq:ae}) at $\gtrsim 2\sigma$ level indicating less dispersion in the relative orientation of PT-PEVs~\citep{aluri2017}.

\begin{table}
\centering
\begin{tabular}{ll}
\hline
Statistic & Multipole(s), $l$ \\
\hline
Power Entropy & 13, 17, 30\\
Aligned with Quadrupole & 3 \\
Aligned modes & (2, 3), (3, 16), (6, 17), (7, 19), (7, 56),\\
              & (8, 10), (10, 18), (11, 35), (13, 15),\\
              & (16, 40), (21, 61), (22, 34), (23, 25),\\
              & (29, 30), (34, 39)\\
\hline
\end{tabular}
\caption{Common set of anomalous multipoles across data sets corresponding to each statistic listed in column 1 are consolidated here, whose $p$-values for a particular statistic were found to be $p\leq0.05$.}
\label{tab:common-modes-pval5pc}
\end{table}

\begin{figure}
\centering
\includegraphics[width=0.98\columnwidth]{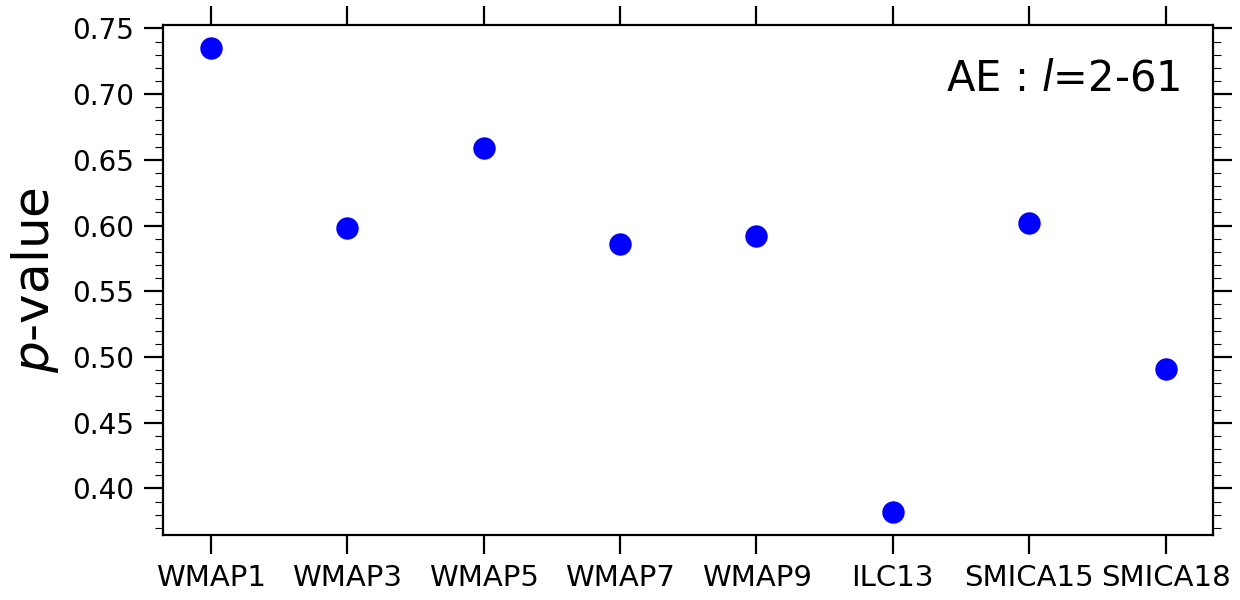}
~
\includegraphics[width=0.98\columnwidth]{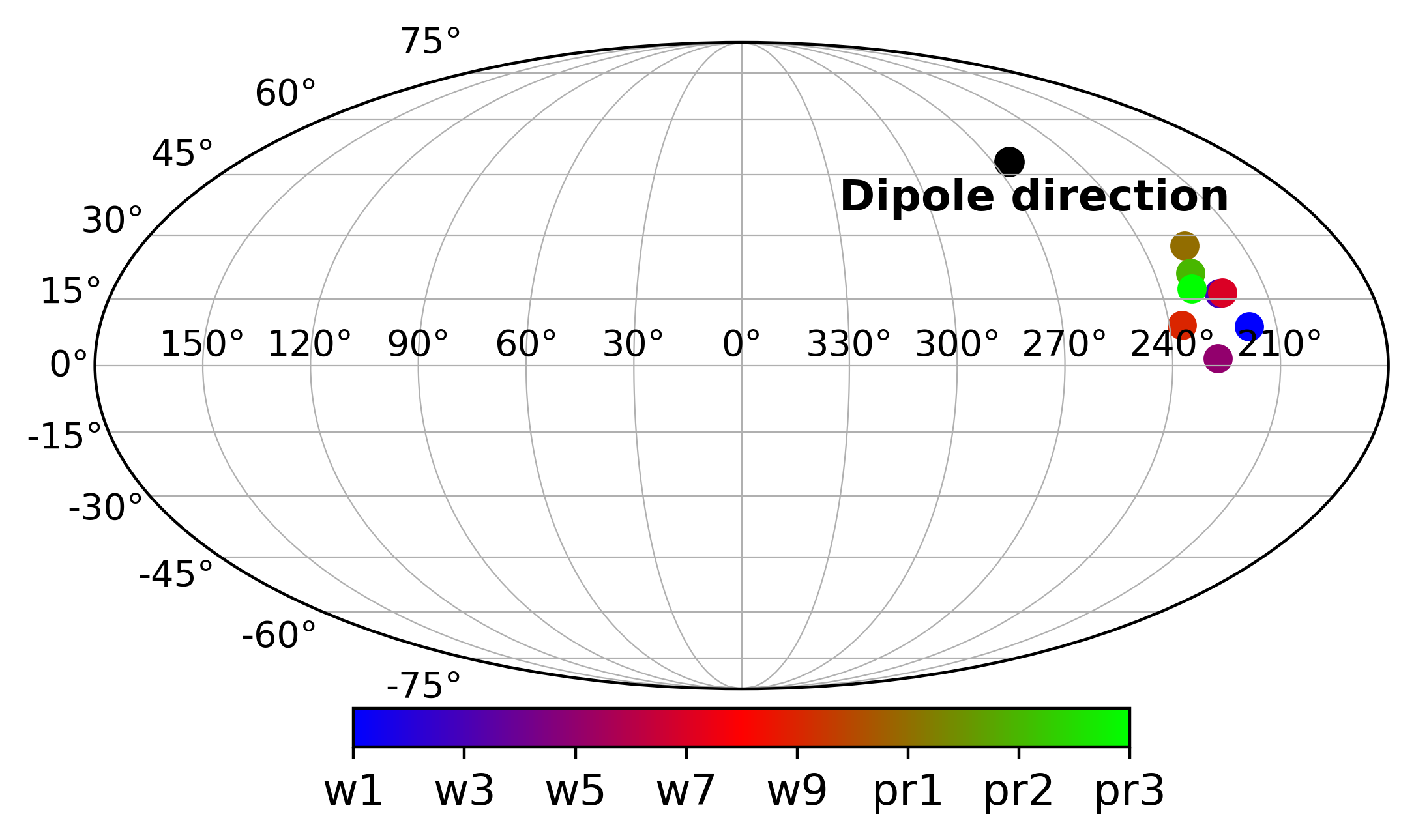}
\caption{
\emph{Top:} $p$-alues of alignment entropy (listed in the order of the tables) are not unusually small.
{\emph Bottom:} Collective alignment axes i.e., $\tilde{\bf f}$ (AT-PEV) from the alignment tensors (equation~\ref{eq:at}) of multipoles $l=[2,61]$ for the 8 data releases.
}
\label{fig:at-plots}
\end{figure}

\subsection{Collective alignments of multipoles}
\label{sec:collective}
The \emph{log}-like statistic employed in Section~\ref{sec:like} revealed that, collectively, the first 60 cosmological modes viz., $l=[2,61]$ are intrinsically anisotropic at greater than 95 per cent confidence level across data releases. Then, it is only natural to query whether there is any coherent alignment among these modes. Here, we test for the presence of any {\it collective} alignments among the first 60 multipoles in CMB maps from the 8 data releases. The test statistics are defined based on the alignment tensor (as per equation~\ref{eq:at}). 

Reviewing this briefly, AT is an entirely directional statistic, which is independent of the eigenvalues of PT. AT is formed from the sum of the outer products of the PEVs over the multipoles $l=2\hdots61$ from each data release. The eigenvector corresponding to the largest eigenvalue of AT (the AT-PEV) probes any {\it collective preferred}  alignment axis for the set of multipoles being studied. AT is isotropic (diagonal with equal entries) in the null model, i.e. the standard cosmological model based on cosmological principle.

AT has its own eigenvalues, which determine alignment entropy by equation~(\ref{eq:ae}). In an isotropic sky, AE is maximal. In that case, no correlation of PT-PEVs would be observed. Indeed, no correlations are observed by far, as is evident from the \emph{top} panel of Fig.~[\ref{fig:at-plots}]. AE from data are comparable to random chance, as indicated by these $p$-values.
The directions of AT-PEVs are shown in the \emph{bottom} panel of the same figure essentially pointing in the same direction with some scatter from all eight data releases. Here, we note that the entropies indicate that the ensembles of PEVs making the AT tend to be isotropic in a statistical sense. However, the AT-PEV is determined by whatever breaks the rotational symmetry of the ensemble. A few well-aligned PEVs can have minimal effects on the entropy of a set of 60 multipoles, while completely determining the AT-PEV.

\subsection{Correlations with CMB dipole, Galactic plane and Ecliptic plane}
\label{sec:algn-special-direc}

\begin{table*}
\centering
\begin{tabular}{c l c c l c c l c c}
\cmidrule(lr){2-4} \cmidrule(lr){5-7} \cmidrule(lr){8-10}
 & \multicolumn{3}{c}{CMB Dipole ($\parallel$)} & \multicolumn{3}{c}{Galactic plane ($\perp$)} & \multicolumn{3}{|c|}{Ecliptic plane ($\perp$)} \\
 \cmidrule(lr){2-4} \cmidrule(lr){5-7} \cmidrule(lr){8-10}
Data set & Multipoles, $l$ & Cumulative & $p$-value & Multipoles, $l$ & Cumulative & $p$-value & Multipoles, $l$ & Cumulative & $p$-value\\
  &  & Probability & of $V$ &  & Probability & of $V$ &  & Probability & of $V$\\
\hline
WILC 1yr   & 2, 21, 41, 42, 61   & 0.180 & 0.771 & 47, 51 & 0.808 & 0.670 & 27, 29, 52, 54 & 0.353  & 0.870 \\
WILC 3yr   & 21, 42, 44, 61      & 0.353 & 0.601 & 51     & 0.954 & 0.254 & 8, 31, 47      & 0.583  & 0.627 \\
WILC 5yr   & 21, 42, 44, 61      & 0.353 & 0.371 & 55     & 0.954 & 0.232 & 8, 26, 31, 47  & 0.353  & 0.927 \\
WILC 7yr   & 21, 36, 42, 44, 61  & 0.180 & 0.733 & 51     & 0.954 & 0.586 & 8, 10          & 0.808  & 0.752 \\
WILC 9yr   & 21, 42, 44, 61      & 0.353 & 0.768 & 57     & 0.954 & 0.797 & 8, 31, 47, 49  & 0.353  & 0.639 \\
PR1 ILC    & 3, 21, 42, 44, 61   & 0.180 & 0.921 & -      & 1     & 0.804 & 18, 47         & 0.808  & 0.863 \\
PR2 \smica\ & 2, 21, 36, 42, 44, 61 & 0.079 & 0.595 & 55  & 0.954 & 0.570 & 2, 8, 47       & 0.583  & 0.915 \\
PR3 \smica\ & 2, 21, 42, 44, 61  & 0.180 & 0.909 & 35     & 0.954 & 0.570 & 2, 8, 10, 47   & 0.353  & 0.830 \\
\hline
\end{tabular}
\caption{List of anomalous multipoles whose alignment measures have a probability of $p\leq0.05$, cumulative binomial probability of having found the observed number of anomalous modes (using $\mathbb{P}=0.05$), and test of random distribution of PEVs using Kuper's `V' statistic with respect to CMB dipole (columns 2–4), Galactic plane (columns 5–7), and ecliptic plane (columns 8–10) when compared with simulations. Alignment statistic `x' and Kuiper statistic `V' are defined in the text.}
\label{tab:algn-special-l1-gal-ecl}
\end{table*}

Finally, we test for alignment of multipoles with some known directions/planes in the sky. Specifically, we test whether the multipole PEVs are spuriously aligned with the CMB dipole direction ($l=1$ mode), the Galactic plane, and the ecliptic plane. We remind that PEVs represent axes and not directed vectors. We use the alignment measure $x:=1-\hat{\lambda}\cdot\tilde{\bf e}_l=1-\cos\alpha_l$, where $\alpha_l=[0^\circ,90^\circ]$, $\hat{\lambda}$ is one of the three known directions/planes in the sky mentioned above, and $\tilde{\bf e}_l$ is the PEV of a multipole `$l$'.
We take the three directions ($\hat{\lambda}$) to be : ${\bf d} = (\ell,b)\approx(264^\circ,48^\circ)$ for the CMB dipole as noted earlier, $(\ell,b)=(0^\circ,90^\circ)$ for the Galactic (north) pole, and $(\ell,b)\approx(276.4^\circ, -29.8^\circ)$ for Ecliptic (south) pole in galactic coordinates\footnote{\url{https://ned.ipac.caltech.edu/coordinate_calculator}}.
We check for the instances when $\{x_{\rm sim}\}$ are smaller than `$x_{\rm obs}$' to test for alignment with CMB kinematic dipole, and when $\{x_{\rm sim}\}$ are larger than `$x_{\rm obs}$' to probe for multipole PEVs lying in Galactic/ecliptic plane. The threshold for significance is $p \leq 0.05$.
 
The results are presented in Table~\ref{tab:algn-special-l1-gal-ecl}. Given that 60 multipoles have been searched, it is not surprising that little of significance is observed. Nevertheless, it is interesting that the multipoles $l=21$, 42, 44 and 61 were found to be consistently well aligned with the CMB dipole direction with $p\leq5$ per cent  in almost all CMB maps.

\section{Summary and Conclusions}

\begin{figure*}
\centering
\includegraphics[width=0.94\columnwidth]{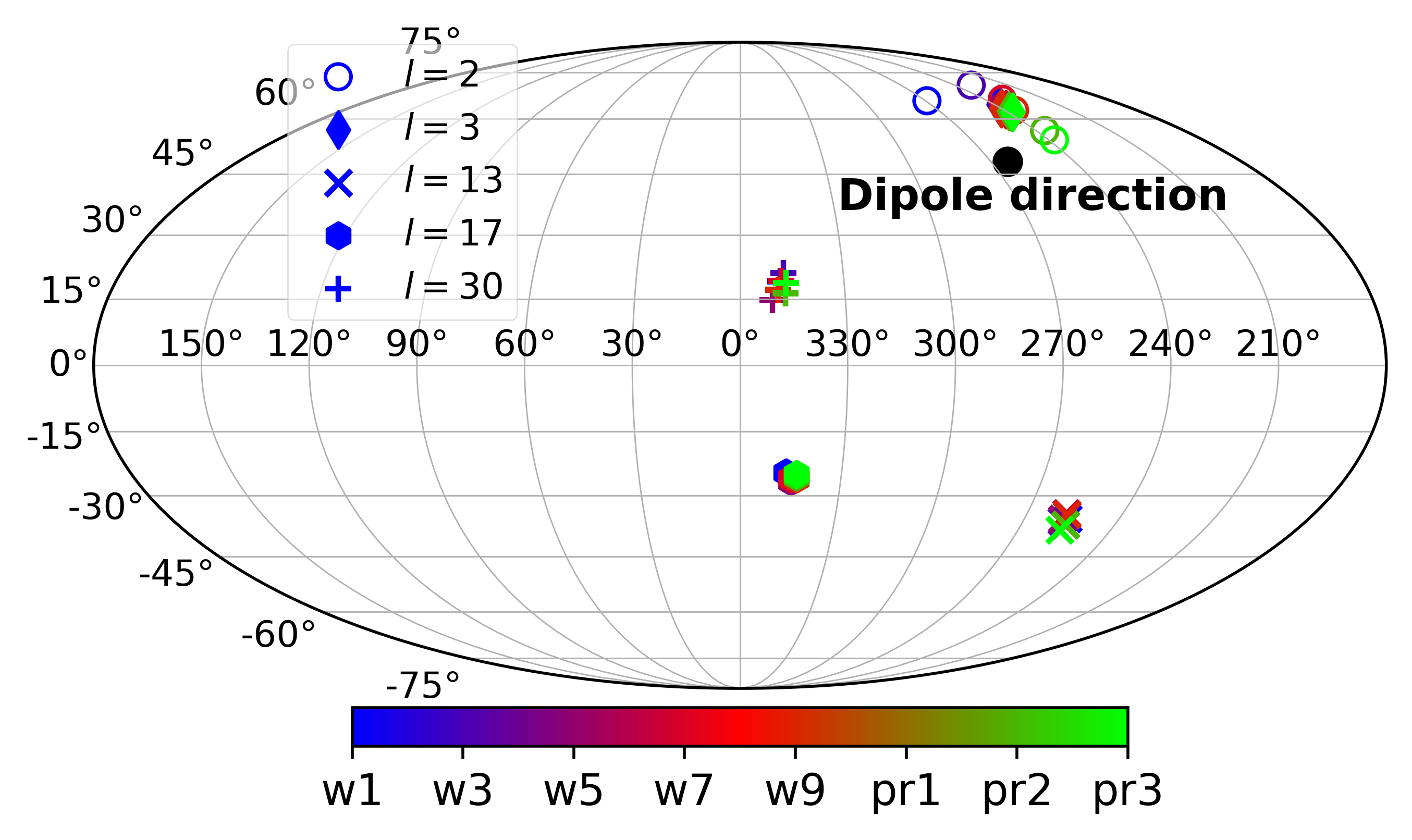}
~
\includegraphics[width=0.94\columnwidth]{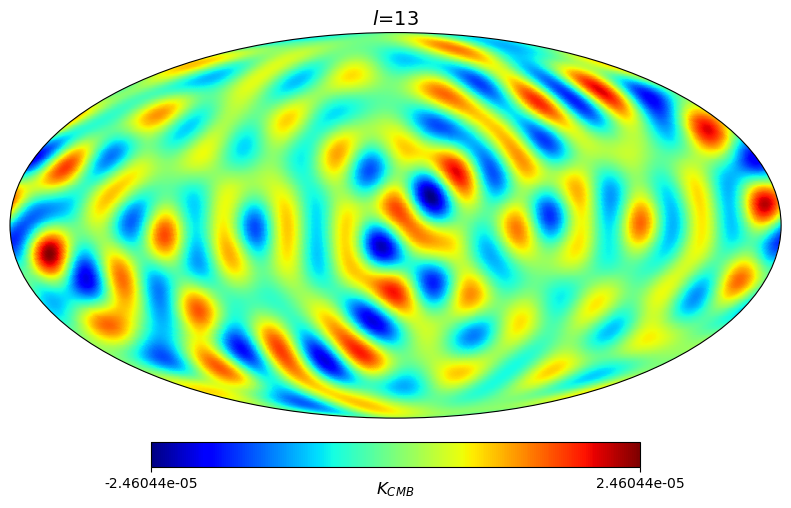}
~
\includegraphics[width=0.94\columnwidth]{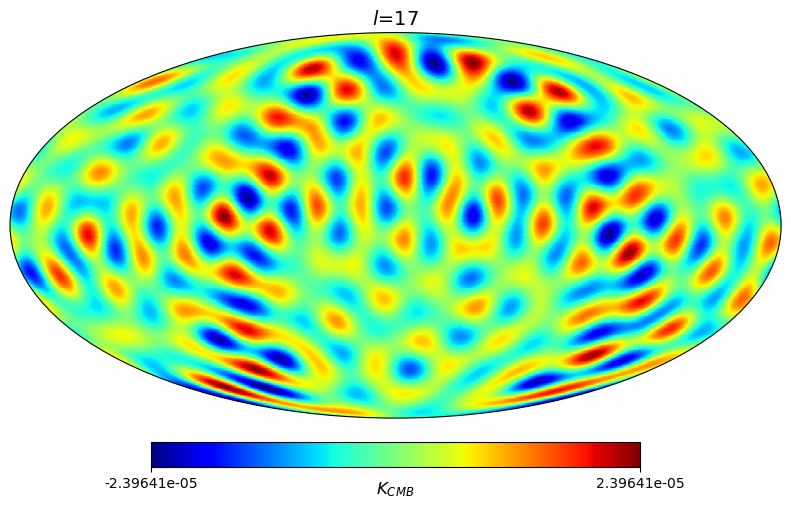}
~
\includegraphics[width=0.94\columnwidth]{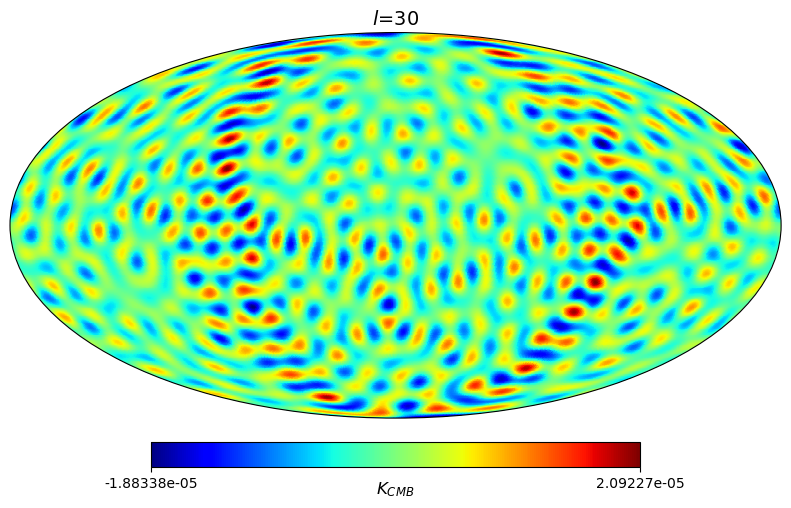}
\caption{\emph{Top :} The anisotropy axis of consistently aligned modes $l=2$, 3 along with CMB dipole ($l=1$) and the PEVs of multipoles $l=13$, 17 and 30 having consistently anomalous PE across all releases. 
\emph{Middle and Bottom :} 
Multipole patterns of $l=13$, 17, and 30 with anomalous PE in galactic coordinates from \planck's 2018 \smica\ CMB map whose random chance occurrence probability is $p\leq 5$ per cent.}
\label{fig:pevl23-anom-pe}
\end{figure*}

We analysed all full-sky CMB maps that are publicly available from NASA’s WMAP’s 1, 3, 5, 7, and 9 yr observational data, as well as CMB maps from ESA’s \planck's 2013, 2015, and 2018 public releases. The eight data releases consist of the ILC method cleaned CMB maps from WMAP’s successive data releases and \planck's nominal (PR1) data, and the \smica\-cleaned CMB maps from \planck's full (PR2) and legacy (PR3) data releases. Our test statistics viz. the \emph{power tensor} (PT) and \emph{alignment tensor} (AT) are rotationally covariant and invariant measures of individual multipoles and collections of multipoles. A statistic is ``anomalous'' if its $p$-value relative to extensive release-by-release simulations is less than 0.05. The effects of conducting multiple searches are controlled by evaluating statistics with a cumulative binomial null distribution and the distribution of log-likelihood statistics.

\subsection{Review of our results}
\begin{itemize}
\item The \emph{power entropy} (PE) of multipoles is consistently anomalous across all releases.
Figure~\ref{fig:pe} shows $p$-values for each `$l$' over all data releases that we studied. The olive green curve in the figure indicates $p\leq 0.05$. Tables~\ref{tab:pe-pval} and \ref{tab:log-like} suggest probabilities of the null model to yield the statistics by chance that are of the order of a few per cent.

In the course of this study, we discovered a bias that underestimates discrepancies from the null, yet which is very common. The bias consists of basing analysis on an arbitrary $p$-value threshold for significance, then generating many trials, and finally not taking into account $p$-values found to be smaller than the threshold. This is a feature of the pass/no pass binomial statistic, which has no information about how far below the $2\sigma$ CL curve the points of Fig.~[\ref{fig:pe}] might be. The first version of the paper was restricted to using strict threshold and fell victim to that bias.
Table~\ref{tab:log-like} was generated using the log-likelihood of the entire set of statistics in the simulated null distribution. This method has no dependence on a $p$-value threshold, and showed that the probabilities of the null to pass the test had been overestimated by the 5 per cent cut-off. (This is not a trivial result: The number of large $p$-values affects the likelihood statistic just as well as the small ones.)  Figure~\ref{fig:PE-PValueCumulative} tells the same story graphically. The cumulative distribution of $p$-values of statistics coming from a distribution that fits the data is a straight line. The distribution of $p$-values compared to an inappropriate distribution deviates from a straight line, with the deviation of Fig.~[\ref{fig:PE-PValueCumulative}] showing that there are too many small $p$-values for the isotropic null distribution to fit the data.

\item Figure~\ref{fig:algn-withl2} and Table~\ref{tab:algn-l2-pval} present test results for alignments of higher multipoles with the quadrupole, $l=2$. The highly significant alignments observed in the CMB map from WMAP's first data release are seen in all releases. However, extending the search over the full range $2\leq l \leq 61$ does not find enough more alignments to indicate a test violating the null hypothesis with high significance. Not unexpectedly, extending the search to all possible pairwise combinations involves so many combinations that nothing of significance is observed, as seen in Table~\ref{tab:algn-ll-pval}. 

\item We also tested for correlation among PT-PEVs with the CMB dipole ($l=1$) direction, the Galactic plane, and the ecliptic plane. While anomalous alignments are found (Table~\ref{tab:algn-special-l1-gal-ecl}), they are distributed across 60 multipoles of the study, which overall suggest no extraordinary correlations.

\item Collective alignments are measured by AT, which is constructed entirely from eigenvectors of PT. There is no importance weighting by eigenvalues in AT. The entropies of AT across releases are unremarkable. The bottom panel of Fig.~[\ref{fig:at-plots}] shows AT-PEVs broadly pointing in the same direction with little scatter.

\item Figure~\ref{fig:pevl23-anom-pe} shows the PEV's of $l=2,\, 3$, the dipole direction, plus the directions of multipoles with anomalously low PE, viz. $l=13$, 17 and 30. It is very remarkable that the PE $p$-values of this set are very anomalous across all releases (Table~\ref{tab:common-modes-pval5pc}). We remind that PE is a measure of the invariants of PT and is completely independent of its eigenvectors. The power entropies of multipoles have no statistical relation to their directions in the isotropic model. PE does not at all suggest the clustering of $l=1,2,3$ axes seen in the upper left panel of Fig.~[\ref{fig:pevl23-anom-pe}].
This was quantified in Sec.~\ref{sec:l123vol} using a volume statistic and found that the CMB dipole, quadrupole, and octopole modes are well aligned at a significance of $2\sigma$ level or better in all releases.

\item Some important comments regarding the methodology and procedures adopted in producing WMAP's ILC-cleaned (WILC) CMB maps and \planck's \smica\-cleaned CMB maps are in order. First, the harmonic space methods are generally more efficient at cleaning raw maps, as they clean foregrounds not only by the level of their spatial distribution (Fig.~[\ref{fig:ilcskyreg}]), but also by angular size i.e., `$l$' (equation~\ref{eq:ilcwghts} and \ref{eq:hilcwghts}). Secondly, a foreground bias correction map is subtracted from WILC maps based on simulations that includes cleaning simulated raw satellite maps with CMB, foregrounds, and appropriate instrument beam and noise effects. This bias correction was shown to improve the alignment between quadrupole and octopole~\citep{aluri2011}, which is not applied to any of \planck's component separated maps including \smica\ CMB map. Further, substantial care is taken to handle systematics in \planck\-produced CMB maps (see for example \cite{santanu2015} and \cite{npipe2020}). Finally, we also note that correcting for (frequency-dependent) kinematic quadrupole was shown to improve the alignment between $l=2,3$ modes~\citep{NotariQuartin2015}. This correction is, however, not applied to any of the CMB maps studied here.

\end{itemize}

\subsection{Conclusions} 

Appendix reports on extensive painstaking simulations, which are heart of our analysis. Here are some observations from our results.

The unexpected alignment seen in WMAP’s first year data release between two of the largest scale multipoles, namely the quadrupole ($l=2$) and octopole ($l=3$) continues to be observed across all releases. This along with the dipole ($l=1$) alignment hints at a preferred direction for our Universe on large scales in the direction of the Virgo cluster. For a long time, the dipole component was assumed to be a kinematic effect of no cosmological significance. It was nearly forgotten that the isotropic cosmological model allows a non-zero dipole fluctuation to exist, which would add linearly to a kinematic component. It is curious that in the null model it would not matter either way whether the dipole is included among the statistics, or omitted. A habit of omitting in ``by convention'' cannot be called objective. The interpretation of the CMB dipole has also become controversial recently in studies of galaxy distributions (see, for example~\cite{Singal2011,Rubart2013,Tiwari2015,Secrest2022}). In summary, the status of the dipole has not been settled, but the anomalous alignment of low-$l$ multipoles cannot be doubted. It is unsurprising that extending the range over 60 multipoles does not turn up much of significance, while we are obliged to report it once we had done it.

Next, we have found strange outcomes we cannot explain. There are three modes $l=13, 17$ and $30$ whose PE is intrinsically anisotropic across all data sets, yet were not reported to be anomalous in earlier studies. The multipole pattern of $l=13$ mode as shown in Fig.~[\ref{fig:pevl23-anom-pe}] reveals that most of its hotspots and cold spots lie along the ecliptic plane. That may suggest a non-cosmological nature. The multipole patterns of $l=17$ and $30$ modes, shown in the same figure, are not self-explanatory, though one might see in them some correlation with our Galactic north polar spur (seen more readily in low-frequency microwave/radio observations). We have no explanation for the rotationally invariant {\it entropy} to select multipoles whose totally independent {\it directional orientations} align as seen.

Finally, there are far too many multipoles with anomalously small PE to be consistent with the null model. The CMB power spectrum has averaged over information in order to make rotationally invariant statistics. Yet the power spectrum is not the {\it only} rotationally invariant quantity available. In everyday terms, the PE is the measure of the ``shape'' of a multipole, visualized as a temperature distribution on a sphere. The most common, most likely PE in the isotropic model, is a shape that is as spherical as possible. A multipole of order `$l$' has $2l+1$ real components, which range from 3 components (the dipole) to 123 components ($l=61$). A typical multipole from the isotropic cosmological model has no discernible shape or orientation, but is more like a riot of random patches. The anomalous number of low-PE shapes observed in the data is more like the orderly patterns seen for single spherical harmonics. PE is a genuine information entropy, as defined by von Neumann and Shannon, and then a genuine probe into the phases of the CMB radiation. The conclusion of our study is that the observed CMB is not as random as the cosmological principle predicts. It would be interesting if these phenomena could be confronted by alternative cosmological models, and more importantly are seen in CMB polarization observations with comparable signal-to-noise ratio in the future.

\section*{Acknowledgements}
Some of the results in the current work were derived using the publicly available {\tt HEALPix}/{\tt Healpy} package~\citep{healpix,healpy}.
This work made use of \texttt{CAMB}\footnote{\url{https://camb.info/}}~\citep{camb1,camb2},
a freely available Boltzmann solver for CMB anisotropies. We acknowledge the use of NASA’s WMAP data from the LAMBDA, part of the High Energy Astrophysics Science Archive Center (HEASARC). HEASARC/LAMBDA is a service of the Astrophysics Science Division at the NASA Goddard Space Flight Center. Part of the results presented here are based on observations obtained with \planck\, an ESA science mission with instruments and contributions directly funded by ESA Member States, NASA, and Canada. We also acknowledge the use of isap software~\citep{isap}.
This research used resources of the National Energy Research Scientific Computing Center (NERSC), a U.S. Department of Energy Office of Science User Facility operated under Contract No. DE-AC02-05CH11231. Further, this work also made use of \texttt{SciPy}\footnote{\url{https://scipy.org}}~\citep{scipy},
\texttt{NumPy}\footnote{\url{https://numpy.org}}~\citep{numpy},
\texttt{Astropy}\footnote{\url{http://www.astropy.org}}~\citep{astropy2013,astropy2018,astropy2022}
and \texttt{matplotlib}\footnote{\url{https://matplotlib.org/stable/index.html}}~\citep{matplotlib}.

\section*{Data availability}
This work made use of the publicly available CMB data from WMAP and \planck\ missions, and other details provided in the accompanying supplementary notes. As such, no new data are generated as part of this work.



\bibliographystyle{mnras}
\bibliography{ref_skp_proj1} 



%

\appendix

\section{Data and simulations}
\label{apdx:data-sim}

\subsection{Data}
\label{apdx:data}
WMAP is a space-based NASA mission, which made observations of the microwave sky in five frequency channels, 23, 33, 41, 61, and 94 GHz. These five frequency channels are also referred to as $K,Ka,Q,V,$ and $W$ bands respectively.
Cleaned CMB maps from these five raw satellite data were obtained using the ILC method where suitable weights were used to remove foregrounds. The observed signal in a frequency channel,
$\nu_i$ (or simply $i$), is supposed to be a linear combination given by,
\begin{equation}
\Delta T_i (\hat{n}) = \Delta T_c(\hat{n}) + F_i (\hat{n}) + N_i (\hat{n})\,,
\end{equation}
where $\Delta T_c(\hat{n})$ is CMB, the cosmic signal, that is independent of frequency (in thermodynamic units),
$F_i(\hat{n})$ represents sum of all foregrounds (galactic/extragalactic astrophysical emission) at frequency `$i$' that are frequency dependent, and
$N_i(\hat{n})$ is the detector noise in channel `$i$'. When the observed sky is digitized using
\healpix\ the sky positions, $\hat{n}$, are equivalently described by pixel indices `$p$' whose pixel centres represent the direction of photons reaching us.

Taking linear combination of raw maps from different frequencies, we get an estimate of the CMB sky to be,
\begin{equation}
\Delta \hat{T}_c (p) = \sum_{i=1}^{n_f} w_i \Delta T_i (p)
   = \Delta T_c(p) \sum_{i=1}^{n_f} w_i + \Delta T_{\rm res}(p)\,,
\end{equation}
where $\Delta T_{\rm res}(p) = \sum_{i=1}^{n_f} w_i [ F_i (p) + N_i (p)]$
and $n_f$ are the total number of frequency channels. Thus, the ILC weights are obtained by minimizing the variance of the cleaned map,
$\Delta \hat{T}_c (\hat{n})$, subject to the constraint that $\sum_i w_i=1$ so that the cosmic CMB signal remains untouched while residual term is minimum or effectively zero. The variance of cleaned CMB map is then given by
\begin{equation}
\sigma^2_c = \langle \Delta \hat{T}_c^2 (p) \rangle - \langle \Delta \hat{T}_c (p) \rangle^2 = 
{\bf w}^T {\bf C} {\bf w}\,,
\end{equation}
where ${\bf w} = (w_1, w_2,\hdots, w_{n_f})^T$ is the $n_f \times 1$ column vector of weights,
$\langle ... \rangle$ represents expectation value of a quantity inside the angular brackets, and
\begin{equation}
C_{ij} = [{\bf C}]_{ij} = \frac{1}{N_{\rm pix}} \sum_{p=1}^{N_{\rm pix}} (\Delta T_i(p)-\overline{\Delta T}_i) (\Delta T_j(p)-\overline{\Delta T}_j)\,,
\end{equation}
is the covariance matrix between different frequency maps and $\overline{\Delta T}_i$ is the
average of all pixels of frequency map `$i$'.

\begin{figure}
\centering
\includegraphics[width=0.45\textwidth]{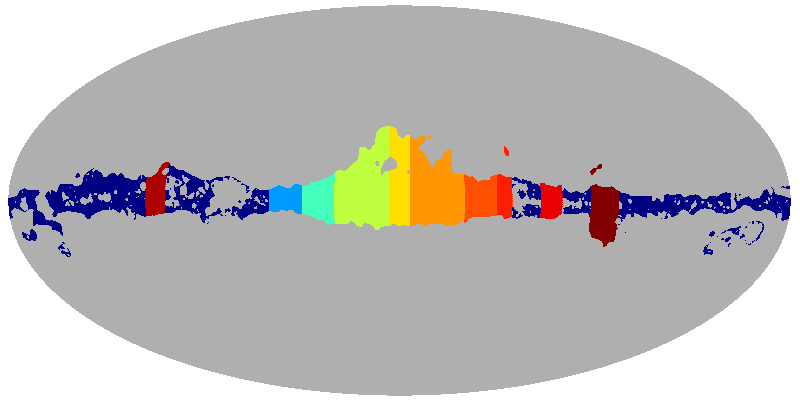}
\caption{Mask from WMAP 9 yr data release depicting the 11 regions used to obtain the composite ILC map. Regions are selected depending on the strength of the foreground signal that varies across the sky (from grey to blue to green to dark red).}
\label{fig:ilcskyreg}
\end{figure}

Since the foreground strength varies across the sky, the ILC procedure is applied by dividing the observed sky into different regions as shown in Fig.~[\ref{fig:ilcskyreg}]. In WMAP’s 1 yr data release, a cleaned CMB map using ILC method was obtained by a non-linear minimization procedure~\citep{wmap1fg}. However \cite{eriksen04} showed that the ILC weights can be obtained analytically using the Lagrange multiplier method as

\begin{equation}
\hat{{\bf w}} = \frac{{\bf C}^{-1}{\bf e}}{{\bf e}^T{\bf C}^{-1}{\bf e}}\,,
\label{eq:ilcwghts}
\end{equation}
where ${\bf e} = (1,1,\hdots,1)^T$ is a $n_f \times 1$ column vector,
and ${\bf \hat{w}}$ are the estimated weights for taking linear combination of raw satellite maps. Cleaned CMB maps thus obtained using the ILC procedure as provided by the WMAP team were used in this study.

To employ real/pixel-space ILC method, all the input raw maps should have the same beam and pixel resolution. So, the raw maps having different beam and pixel resolution were repixelized so that they have a common beam resolution given by a Gaussian beam of $FWHM=1^\circ$
at \nside=512 following,
\begin{equation}
a_{lm}^{\rm out} = \frac{b_l^{\rm out}p_l^{\rm out}}{b_l^{\rm in}p_l^{\rm in}}a_{lm}^{\rm in}\,,
\label{eq:repixel}
\end{equation}
in harmonic space, where $b_l$ is the circularized beam transfer function and
$p_l$ being the pixel window function corresponding to the \nside\ of a \healpix\
digitized map.
Here $b_l^{\rm in}=b_l^i$ where $i=K,Ka,Q,V,W$ bands of WMAP and $b_l^{\rm out}=b_l^{1^\circ}$
i.e., a one degree FWHM Gaussian beam transfer function. These $a_{lm}^{\rm out}$ are then
converted into a map using \healpix.

\begin{table*}
\centering
\begin{tabular}{ c c c c c c c c }
\hline
Region & 30 & 44 & 70 & 100 & 143 & 217 & 353 \\
\hline\\
 0 & 0.0503858 & -0.0776323 & -1.1811517 & 1.5601608 & 1.5752080 & -0.9763756 & 0.0494050 \\
 1 & 0.0454710 & -0.1282175 & -0.6392035 & 1.3575588 & 1.2143440 & -0.9020380 & 0.0520853 \\
 2 & 0.0013979 & -0.2325734 &  0.0506873 & 0.8282015 & 0.9601726 & -0.6438950 & 0.0360092 \\
 3 & 0.0319588 & -0.2345117 & -0.0626022 & 0.7419377 & 0.9278901 & -0.4114967 & 0.0068239 \\
 4 & 0.0209488 & -0.1992118 & -0.2429189 & 0.6855018 & 1.1617237 & -0.4334980 & 0.0074544 \\
 5 & -0.0180088 & -0.3461348 &  1.0192336 & -0.2018088 &  0.3969976 & 0.1843588 & -0.0346377 \\
 6 & 0.0668860 & -0.0873926 & -0.9815614 & 0.9276151 & 1.6082933 & -0.5349676 & 0.0011272 \\
 7 & 0.0644619 & -0.0083785 & -1.2588580 & 1.0754668 & 1.7835900 & -0.6665963 & 0.0103141 \\
 8 & 0.2388520 & -0.7999371 & -0.3497620 & 1.1749154 & 1.4655975 & -0.7559560 & 0.0262902 \\
 9 & -0.0311495 & -0.1769642 &  0.6870286 & -0.6905072 & 0.5803113 &  0.7429941 & -0.1117130 \\
10 & -0.0000140 &  0.0377577 & -1.0804496 & 2.0914197 & 1.3302067 & -1.4947717 & 0.1158512 \\
11 & 0.0335069 & -0.1065167 & -0.8299705 & 1.1446248 & 1.5446750 & -0.8249751 & 0.0386556 \\
\hline
\end{tabular}
\caption{ILC weights corresponding to the 12 regions (indexed 0 to 11) defined by the WMAP 9 yr regions’ mask shown in Fig.~[\ref{fig:ilcskyreg}] to derive a cleaned CMB map from \planck\ PR1 data. All the frequency bands labeled 30, 44, $\hdots$, 353 are in units of `GHz'.}
\label{tab:wghts-pr1ilc}
\end{table*}

We also made use of ESA's \planck\ satellite data as provided by \planck\
collaboration through their 2015 and 2018 data releases (for which complementary simulations are available). Particularly, we made use of the \smica\ cleaned CMB maps in this work. They are available at a much higher resolution of \healpix\ \nside=2048. So, these are downgraded to get \smica\ CMB maps at \nside=512 smoothed to have a Gaussian beam $FWHM=1^\circ$, just like WMAP's ILC CMB maps following equation~(\ref{eq:repixel}). We note that \smica\ foreground cleaning procedure is also an ILC method, but performed in harmonic
space using $a_{lm}^i$ i.e., spherical harmonic coefficients of raw satellite maps from different frequencies in which \planck\ made microwave sky observations viz., 30, 44, 70, 100,
143, 217, 353, 545, 857~GHz. The weights are computed using the formula,
\begin{equation}
\hat{{\bf w}}(l) = \frac{{\bf C}^{-1}_l{\bf e}}{{\bf e}^T{\bf C}^{-1}_l{\bf e}}\,,
\label{eq:hilcwghts}
\end{equation}
similar to equation~(\ref{eq:ilcwghts}),
where $C_{ij}(l) = \sum_{m=-l}^l a_{lm}^i a^{j*}_{lm}$.
Specifically, weights for the \smica\ procedure are obtained by using fitted elements of the
cross frequency channel covariance matrix $C_{ij}(l)$ as a function of `$l$' via a minimization procedure. For more details, the reader may consult~\cite{smica2008}.

Now in order to use \planck\ 2013 public release 1 (PR1) data, the complementary simulations are
no longer available with the release of \planck\ PR2 and PR3 data sets. However, we can still make use of this data set by employing a cleaning procedure, and then processing the simulations in the same way. Here, we adopt the pixel-space ILC method, described above, to clean the \planck\ 2013 raw maps. But we use only the observed raw satellite maps in 30, 44, 70, 100, 143, 217, and 353~GHz frequency bands to produce cleaned CMB
sky, that is referred to as \emph{PR1 ILC}. To clean the satellite maps, we used the
WMAP 9yr region masks (shown in Fig.~[\ref{fig:ilcskyreg}]) for iterative cleaning of
microwave sky. All these frequency maps are first downgraded to \nside=512 and smoothed
to have a Gaussian beam of $FWHM=1^\circ$, following equation~(\ref{eq:repixel}).
The ILC weights thus obtained are given in Table~\ref{tab:wghts-pr1ilc}.

\begin{figure}
\centering
\includegraphics[width=0.95\columnwidth]{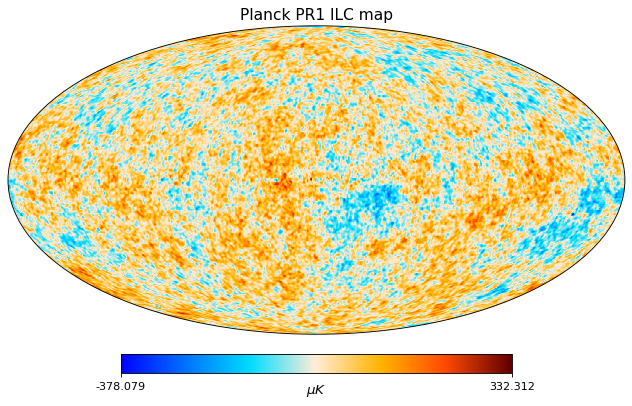}
~
\includegraphics[width=0.95\columnwidth]{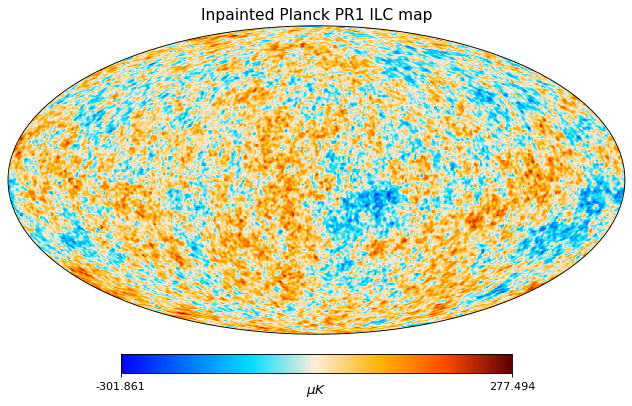}
\caption{\emph{Top:} Cleaned CMB map obtained from \planck\ PR1 data using 30-353~GHz
	raw frequency maps, ILC method.
	\emph{Bottom:} 
	 Same map inpainted using \isap\ with the mask depicted in Fig.~[\ref{fig:our-mask}].}
\label{fig:pr1-ilc}
\end{figure}

The cleaned CMB map obtained using ILC method on \planck\ PR1 data, specifically using
30 to 353~GHz channel raw maps and the corresponding inpainted map thus obtained
are shown in Fig.~[\ref{fig:pr1-ilc}].

\subsection{Simulations}
\label{apdx:sim}
Frequency-specific mock maps for WMAP data are generated as follows. A random realization of the
CMB sky is created at \nside=512 using the fiducial theoretical power spectrum ($C_l^{th}$)
from that release. It is then smoothed with the frequency-specific beam transfer function, $b_l^i$, and is then added with frequency-specific noise. WMAP frequency-specific noise maps are generated as Gaussian noise in each pixel `p' using the noise rms of each channel ($\sigma_0^i$), and the $N_{\rm obs}^i (p)$ map that gives the effective number of observations per pixel in a frequency band `$i$'.
Zero mean, unit variance Gaussian random number arrays of the same dimension as the number of pixels in frequency-specific maps ($N_{\rm pix}=12\times N_{\rm side}^2$)
are generated and then multiplied by the noise rms map per pixel given by $\sigma_i (p) = \sigma_0^i/\sqrt(N_{\rm obs}(p))$ to create a band-specific noise simulation.
It is then added to the channel-specific beam-smoothed CMB realization. Now, all these noisy CMB maps are repixelized following  equation~(\ref{eq:repixel}) so that they have a
common \nside=512 and beam resolution due to a Gaussian beam of $FWHM=1^\circ$.
The noise rms, $\sigma_0(i)$, maps of effective number of observations per pixel,
$N_{\rm obs}^i (p)$, and the WMAP frequency-specific beam transfer functions, $b_l^i$,
are provided as part of each WMAP data release at NASA’s LAMBDA webpage.

Once the smoothed noisy frequency-specific CMB maps are obtained, they are combined using the same region-specific ILC weights obtained from data. Thus, we generate 1000 mock WILC maps corresponding to each data release from WMAP.

Simulated \smica\ CMB maps from \planck\ satellite are provided as part of each data
release by the \planck\ collaboration by processing the FFP simulations in the same way as \smica\ CMB map is obtained from observational data.
\smica\ map is provided at a higher resolution of \nside=2048 with a Gaussian
beam smoothing of $FWHM=5'$. Hence these maps are also repixelized to \nside=512 and whose beam window function is given by a Gaussian with $FWHM=1^\circ$ following equation~(\ref{eq:repixel}) (by taking $b_l^{\rm in}=b_l^{5'}$, $b_l^{\rm out}=b_l^{1^\circ}$, $p_l^{\rm in}=p_l^{2048}$ and $p_l^{\rm out}=p_l^{512}$) similar to WILC maps.

Now to generate mock CMB maps for \planck's PR1 data, the ILC weights used to clean
the raw satellite maps from \planck\ PR1 are applied to the simulated FFP frequency realizations, iteratively, after processing them similar to data.
We recall that \planck\-provided raw maps are available at \nside=1024 for 30, 44, and 70~GHz
bands (LFI bands) and at \nside=2048 for 100, 143, 217, 353, 545 and 857~GHz bands (HFI bands).
So, they are processed to bring all of them to a common beam and pixel resolution of $1^\circ$ FWHM Gaussian beam
and \nside=512 respectively. Thus, we generated 1000 mock PR1 ILC maps from the corresponding frequency-specific FFP realizations. Note that we did not use the last two high-frequency bands in obtaining \planck\ PR1 ILC CMB map.

We further note that WILC maps had a residual foreground bias correction applied based on simulated foreground cleaned CMB maps~(\citep{wmap3temp}). However, such a bias correction procedure was not employed on the CMB maps recovered using any of the four component separation methods (in real or harmonic space) employed by the \planck\ collaboration. So, we also did not apply any such corrections to PR1 ILC CMB map.

\section{Masks used}
\label{apdx:mask}
The presence of residual foregrounds in the \emph{cleaned} CMB maps can bias our inferences. Thus, such regions of the sky are masked and inpainted using the
\isap\ package. However, there will also be a bias introduced in the inpainted CMB maps if large fractions of the sky have to be inpainted. Since we are interested in CMB multipoles $l \leq 61$, we use a single mask that has largely contiguous sky regions, but also masks extended regions with potential foreground residuals. It is obtained by combining WMAP’s 9 yr Kp8 temperature cleaning mask and
\planck's 2018 common inpainting mask for temperature analysis as described below. WMAP’s 9 yr Kp8 mask that is available at \healpix\ \nside=512 is upgraded to
\nside=2048 after removing small point sources from it. Then, it is combined (multiplied) with \planck\ 2018 common inpainting mask provided for intensity data. The combined mask thus obtained is then convolved with a Gaussian beam of $FWHM=1^\circ$
while deconvolving with a Gaussian beam of $FWHM=5'$, and repixelized at \nside=512.
This combined smoothed mask is then thresholded such that those pixels that have a value
$\geq0.8$ are set to `$1$' and rest are set to `$0$'. Mask thus obtained has a sky fraction
of \fsky$\approx 0.929$ which was shown in Fig.~[\ref{fig:our-mask}].

The data as well as simulated CMB maps corresponding to various data releases from WMAP and \planck\ missions are all inpainted using this mask.

\section{Smoothed $K$- and $Ka$-band maps from WMAP first and three year data release}
\label{apdx:wmap1kmap}
There are few details to be noted in the generation of mock WILC CMB maps described in the previous section, and our observations while carrying out this analysis.

Since WMAP's $K$ and $Ka$ frequency channel detectors are of low resolution, the effective circularized beam transfer functions ($b_l^i$) of these two bands were determined up to $l_{max}=750$ and $850$ respectively, by the WMAP team. However, beam transfer functions that are provided cannot be used as such, since
$K$-band $b_l$ from first year data are positive only up to $l=603$, and in three year data
release $K$- and $Ka$-band $b_l$ are positive up to only $l=718$ and $845$ respectively.
(Here, we also note that WMAP's first year $Q$-band $b_l$ are provided only up to
$l_{max}=1000$. Since the maps are being synthasized at \nside=512, in general one uses
mode information up to $l_{max}=2\times N_{side}$ when employing \healpix\ i.e., $l_{max}=1024$ in our case.)

\begin{figure}
    \centering
        \includegraphics[width=0.95\columnwidth]{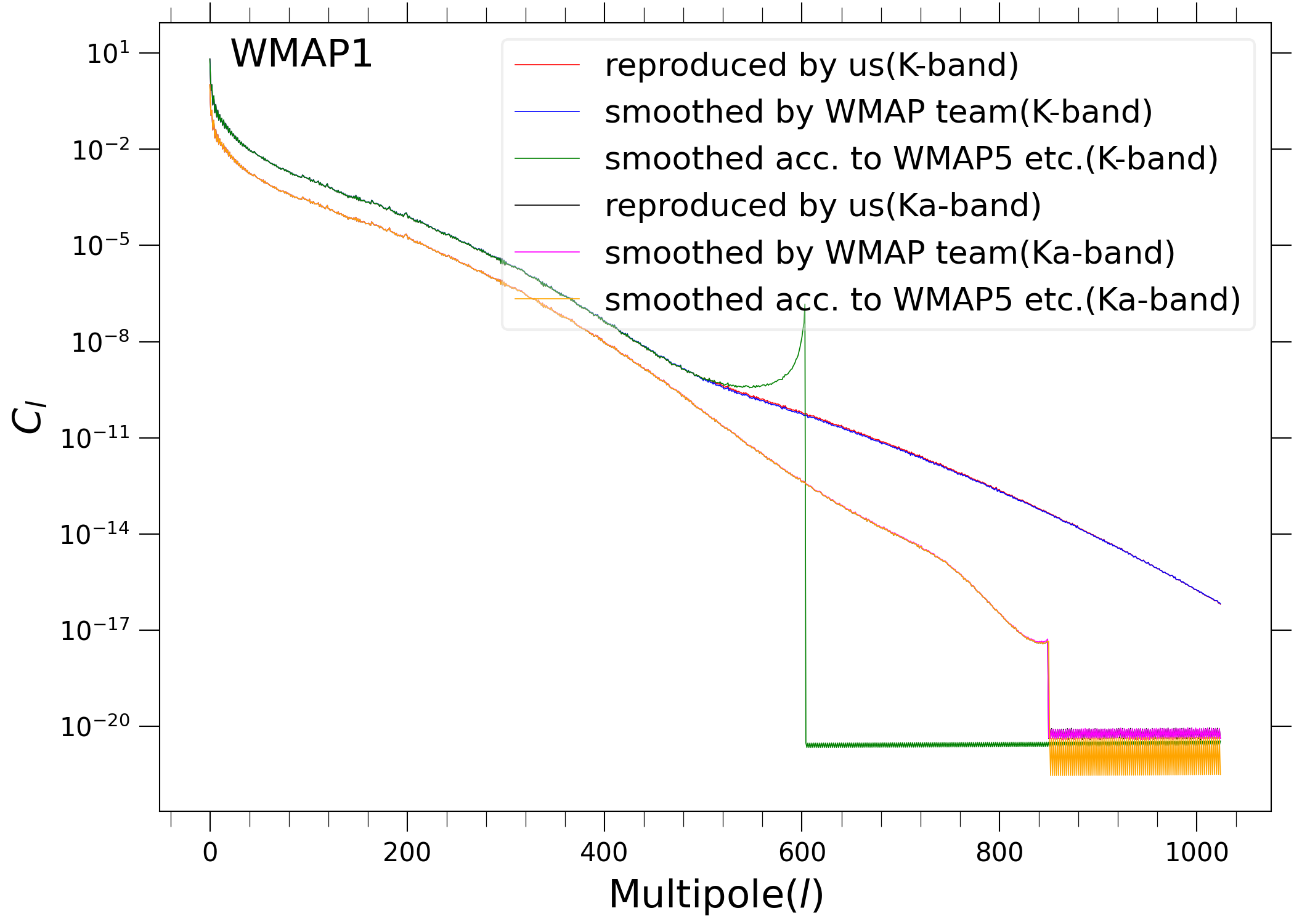}\\
        \includegraphics[width=0.95\columnwidth]{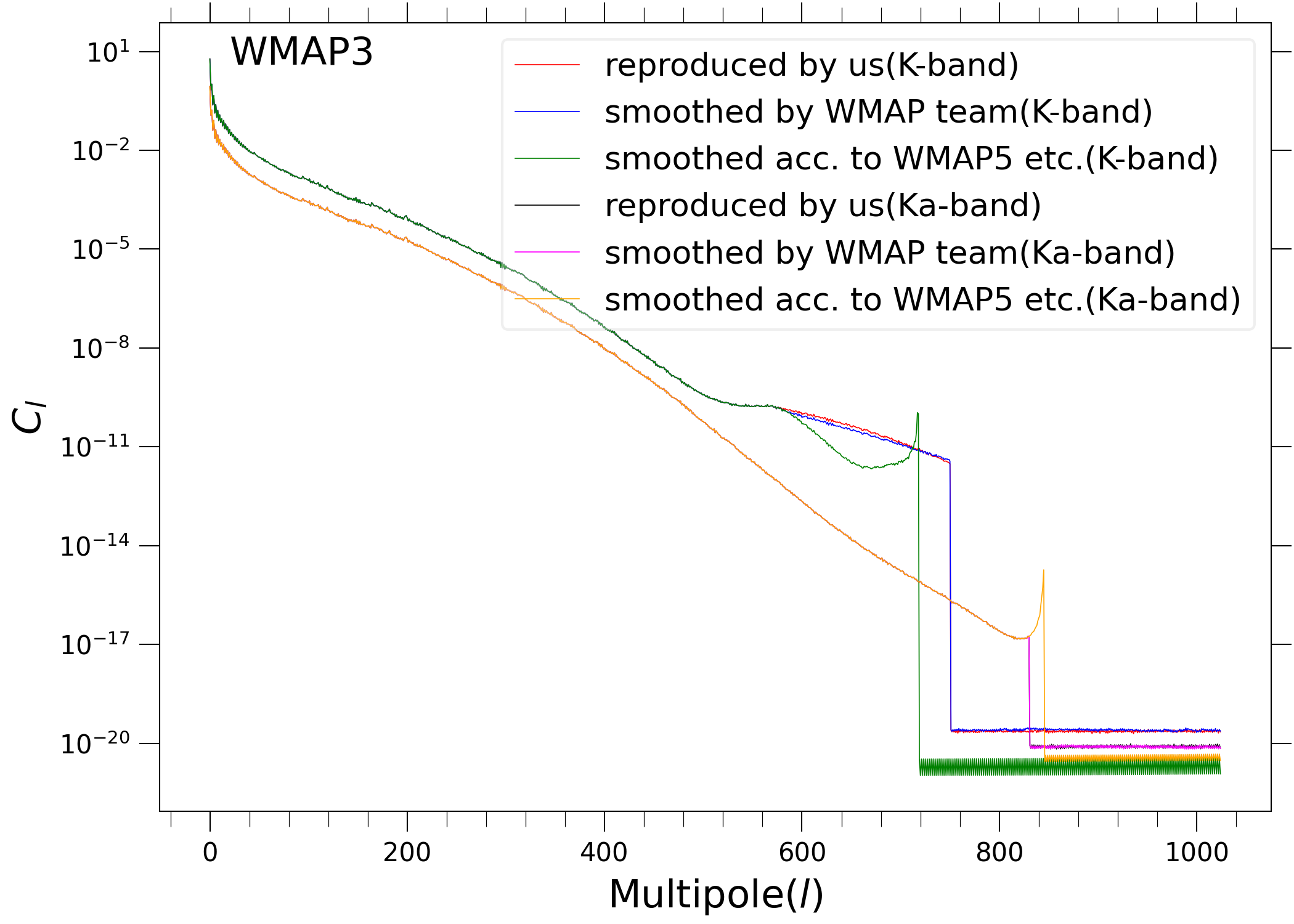}
	\caption{Power spectra of raw satellite maps from WMAP's $K$ and $Ka$ bands provided by
			WMAP team as part of WMAP's first and three year data products, and as obtained
			by us following equation~(\ref{eq:repixel}).}
\label{fig:beam-wmap1and3-k-ka}
\end{figure}

In order to perform ILC cleaning on raw satellite data, the individual frequency-specific maps from WMAP are synthesized at the same pixel resolution of \nside=512 and a common beam smoothing given by a Gaussian kernel of $FWHM=1^\circ$ (which is slightly greater
than the resolution of WMAP's $K$-band detector's effective Gaussian width).
Since the input and output maps are at \nside=512 in
this case, $p_l^{\rm in}=p_l^{\rm out}$. The effective beam smoothing
applied to each frequency map is then $b_l^{i,{\rm eff}}=b_l^{1^\circ}/b_l^i$. Thus in obtaining
$1^\circ$ Gaussian beam smoothed raw maps, only the positive circularized $b_l^i$'s of
$K$ and $Ka$ bands were used in computing $b_l^{i,{\rm eff}}$, and the rest were set to zero.

However, this procedure seems to have been used from WMAP 5 yr data analysis onwards uniformly for processing all frequency maps, but not in 1 and 3 yr WMAP data analysis, especially $K$- and $Ka$-band maps.
Our observations are presented in Fig.~[\ref{fig:beam-wmap1and3-k-ka}].
In the \emph{top} panel, power spectrum plots of WMAP's first year $1^\circ$ Gaussian beam
smoothed $K$- and $Ka$-band maps as available from NASA's LAMBDA site and their power spectrum
as obtained by us following equation~(\ref{eq:repixel}) are shown.
Also shown are the power spectra of $K$- and $Ka$-band maps
that are obtained using respective $b_l^i$ as given, but only up to some low multipole
(i.e., not all positive $b_l^i$ were used), beyond which extrapolated $b_l^i$ are employed.
These extrapolated $b_l^i$ were obtained by us, through trial and error, such that we are able to match the power spectra of smoothed $K$- and $Ka$-band maps thus obtained following
equation~(\ref{eq:repixel}), with the power spectra of smoothed $K$- and $Ka$-band maps as provided by WMAP team.

One readily notices that there is an artificial rise in the power at high multipoles when one uses equation equation~(\ref{eq:repixel}) utilizing all positive $b_l^i$ of $K$- and $Ka$-band maps as given.
This could be the reason for using extrapolated $b_l^i$ by WMAP team for these two band maps
in derving $1^\circ$ Gaussian beam smoothed raw maps that are eventually used in obtaining a cleaned CMB map employing ILC method. However, these $b_l^{i,{\rm eff}}$ o get smoothed raw maps are not made publicly available to be able to appropriately simulate ILC-like maps from WMAP.

The \emph{bottom} panel of Fig.~[\ref{fig:beam-wmap1and3-k-ka}], is same as the \emph{top}
panel but corresponds to WMAP three year $K$- and $Ka$-band maps.  We note that the rise in smoothed raw maps’ power is seen in $K$-band only in first year, and in both $K$- and $Ka$-band map powers in 3 yr data owing to difference in fitted circularized beam transfer functions of corresponding detectors. These combination $b_l^i$
containing partly the circularized $b_l^i$ of $K$ and $Ka$ bands as provided and the extrapolated beam transfer function found by us (through trial and error) were used to simulate smoothed noisy frequency-specific CMB maps to which ILC weights were applied in generating mock ILC-cleaned CMB maps.

We were able to match the power spectrum of WMAP’s first year $1^\circ$ beam smoothed $K$-band raw map, as provided,
with the WMAP 1yr $K$-band map's power spectrum by using $b_l^{K}$ up to $l=551$ as given, and
fitting them using \texttt{UnivariateSpline} method of \texttt{SciPy}
(a \texttt{Python} software library) choosing the degree of spline to be $k=2$
(i.e., using a quadratic spline) to obtain extrapolated $b_l^K$ up to $l=1024$.
Then the extrapolated and given beam transfer functions of $K$-band are combined at $l=508$.
Similarly, for WMAP's 3yr $K$ band, we found (by trial and error) that \texttt{interp1d}
method of \texttt{SciPy} with the paramter choice \texttt{kind=`slinear'} (first order/linear spline) gives extrapolated $b_l$ from $l=570$ onwards such that the combined $b_l$ result in a raw map power spectrum following equation~(\ref{eq:repixel}) that matches with the smoothed raw maps' $C_l$ from WMAP three year data. Thus combined $b_l$ up to a maximum multipole of $l=750$ were
used for obtaining smoothed $K$ band map, and up to $l=830$ were used to get $Ka$ band smoothed map.

\section{Smoothed LFI maps from \planck\ PR1 data}
\label{apdx:plkpr1lfi}
We discussed the problem of appropriately simulating $1^\circ$ Gaussian beam smoothed maps
corresponding to $K$ and $Ka$ bands that match with WMAP's 1yr and 3yr smoothed $K$- and $Ka$-band data maps as provided by WMAP team in the previous section.

\begin{figure}
\centering
\includegraphics[width=0.95\columnwidth]{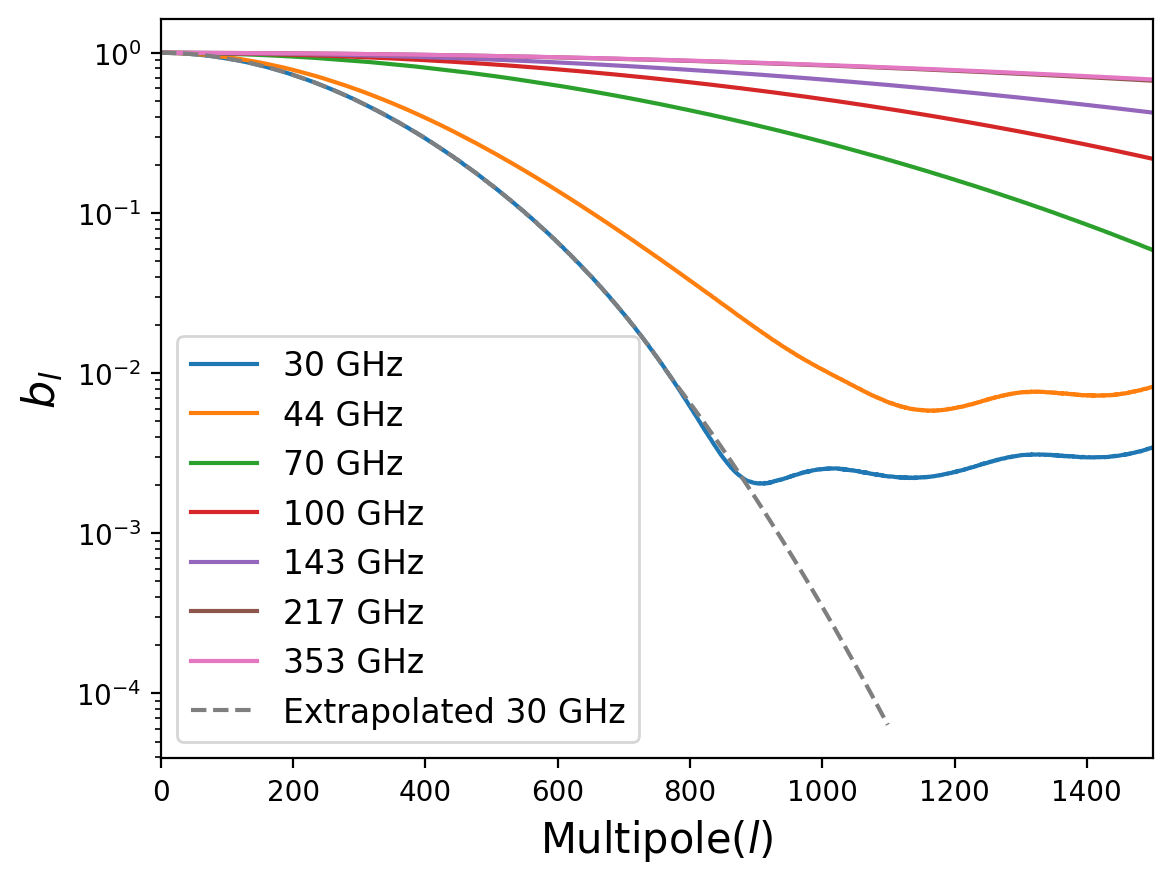}
\caption{
Circular beam transfer functions ($b_l$) of all frequency maps that we used from \planck\ PR1 data.
One can easily notice the unphysical upward trend of the 30~GHz channel’s effective circular $b_l$ as provided by \planck\ collaboration.
Similar trend for 44~GHz is not a concern for our present study.
}
\label{fig:planck-pr1-bl}
\end{figure}

We are faced with a similar problem in trying to employ real space ILC method with \planck\ PR1 low-frequency maps, specifically with the 30 GHz channel. Since we also chose to repixelize all the maps being used in this study to \nside=512 having a beam smoothing given by a Gaussian kernel with $FWHM=1^\circ$, \planck\ 30~GHz channel poses a similar problem. The circular beam transfer functions of all frequency maps from \planck\ 2013
data release are shown in Fig.~[\ref{fig:planck-pr1-bl}].

It is clear from the figure that beam transfer functions of \planck's 30~GHz frequency map from PR1 as provided are not usable beyond $l \sim 850$ or so after which it shows an unphysical upward trend. Hence, we adopt the same strategy as used for WMAP’s $K$- and $Ka$-band maps from the 1 and 3 yr data releases described in the preceding section.

We fit the \planck\ PR1 30~GHz channel's $b_l$ as given for multipoles up to $l=860$.
using a spline. Then, the beam transfer functions as provided up to $l=773$ are appended with the extrapolated
$b_l$ obtained from the fit, where the two cross each other. The extrapolated $b_l$ were obtained by a fit to the given $b_l$ for 30~GHz detector using \texttt{UnivariateSpline} functionality of \texttt{SciPy}, a \texttt{Python} software library, with the parameter $k=2$ i.e., a quadratic spline. The effective (extrapolated) $b_l$ are generate up to $l=1100$ and following equation~(\ref{eq:repixel}) \planck\ PR1 30~GHz map was synthesized at \nside=512 having $1^\circ$ Gaussian beam smoothing with mode information up to $l_{max}=1024$.

One can also see from Fig.~[\ref{fig:planck-pr1-bl}] that the 40~GHz detector beam transfer function also shows a similar unphysical upward trend, but at a much higher multipole value. So, there was no need for such a procedure to repixelize frequency maps other than 30~GHz channel map at \nside=512. The beam transfer functions are well behaved up to $l=1024$ for rest of the frequency maps.

We also note that this can be addressed in two other ways. As in WMAP 5 yr and later data releases, $a_{lm}^i$'s of frequency maps in generating the smoothed maps can be set to zero for modes beyond that multipole for which $b_l$ are not provided or are not physically acceptable. A better way to address this is to use harmonic space ILC using only those modes ($a_{lm}^i$) from frequency channels where $b_l$ are available/acceptable to deconvolve the observed map for detector beam smoothing effects (see for example~\cite{tegmark03}).


\bsp	
\label{lastpage}
\end{document}